\documentclass[reqno,12pt]{amsart}
\usepackage{ifthen}
\usepackage{natbib}
\bibpunct{[}{]}{,}{a}{,}{,}
\usepackage[pdftex]{graphicx}
\DeclareGraphicsExtensions{.pdf}

\newcommand{\version}{journal}

\ifthenelse{\equal{\version}{journal}}%
{\message{
--- journal version ---
}}%
{\message{
--- personal version ---
}
}

\usepackage{amssymb}
\usepackage{amsmath}
\usepackage{amsfonts}
\usepackage{latexsym}

\numberwithin{equation}{section}
\numberwithin{table}{section}
\numberwithin{figure}{section}
\newtheorem{theorem}{Theorem}[section]

\newtheorem{lemma}[theorem]{Lemma}

\newtheorem{example}[theorem]{Example}

\newcommand{\boldc}{{\mathbf c}}

\newcommand{\bbR}{{\mathbb{R}}}
\newcommand{\R}{{\mathbb{R}}}
\newcommand{\C}{{\mathbb{C}}}

\newcommand{\Z}{{\mathbb{Z}}}
\newcommand{\boldm}{{\boldsymbol{m}}}

\newcommand{\boldy}{{\boldsymbol{y}}}

\newcommand{\boldPhi}{{\boldsymbol{\Phi}}}
\newcommand{\boldphi}{{\boldsymbol{\phi}}}

\newcommand{\boldpsi}{{\boldsymbol{\psi}}}

\newcommand{\<}{\langle}
\renewcommand{\>}{\rangle}
\newcommand{\binomial}[2]{\begin{pmatrix}#1\\#2\end{pmatrix}} 

\begin{document}

\title[Two-direction multiwavelet point values and derivatives]
{Point values and normalization of two-direction multiwavelets and their derivatives}

\author{Fritz Keinert and Soon-Geol Kwon$^{*,1}$}

\thanks{$^{*}$Corresponding author}
\thanks{${}^{1}$This paper was supported by (in part) Sunchon National University Research Fund in 2012.}

\address{Fritz Keinert \\
Department of Mathematics, Iowa State University, Ames, IA}
\email{keinert@iastate.edu}
\address{Soon-Geol Kwon \\
Department of Mathematics Education, Sunchon National University,
Sunchon, 540-742, Korea}
\email{sgkwon@sunchon.ac.kr}

\begin{abstract}
Two-direction multiscaling functions $\boldsymbol{\phi}$ and two-direction multiwavelets
$\boldsymbol{\psi}$ associated with $\boldsymbol{\phi}$ are more general and more flexible setting
than one-direction multiscaling functions and multiwavelets.
%
In this paper, we investigate how to find and normalize point values and
those of derivatives of the two-direction multiscaling functions $\boldsymbol{\phi}$
and multiwavelets $\boldsymbol{\psi}$. 
For finding point values, we investigate the eigenvalue approach.
For normalization, we investigate the normalizing conditions for them by normalizing
the zeroth continuous moment of $\boldsymbol{\phi}$.
Examples for illustrating the general theory are given.
\end{abstract}




\maketitle


\section{Introduction}\label{introduction}

Two-direction multiscaling functions $\boldphi$ and two-direction multiwavelets
$\boldpsi$ associated with $\boldphi$, which are more general setting than
one-direction multiscaling functions and multiwavelets, 
have been investigated in a few papers in the literature, for example, see
~\citep{DuYuan2010,Kwon2011c,Kwon2012,LvWang2009,Morawiec2009,YangLi2007AMC,YangLi2007SciChina,YangXie2008}.
The two-direction setting is more flexible than one-direction setting.

In multiwavelet theory, computation of point values and those of derivatives for the
one- or two-direction multiscaling functions and multiwavelets associated with
multiscaling functions is important.
Point values of multiscaling functions and multiwavelets are needed to reconstruct
a function from its expansion coefficients.
Point values and those of derivatives for the one- or orthogonal two-direction
multiscaling functions and multiwavelets associated with multiscaling functions
are necessary in many applications such as solution of differential equations,
signal processing, and image processing.

In one-direction multiwavelet theory, two approaches,
called the cascade algorithm approach and the eigenvalue approach,
are known for finding point values and those of derivatives of the multiscaling
functions and multiwavelets (see~\citep{daubechies88,daubechies92,Keinert2004}).

The cascade algorithm approach computes approximate point values and those of
derivatives of the multiscaling functions and multiwavelets.
The eigenvalue approach computes exact point values and
those of derivatives of the multiscaling functions and multiwavelets, but may fail
in some cases.

Point values and those of derivatives of the multiscaling functions and
multiwavelets can all be computed from the recursion coefficients.

Both the cascade algorithm approach and the eigenvalue approach can also
be considered for finding point values and those of derivatives of the two-direction
multiscaling functions $\boldphi$ and multiwavelets $\boldpsi$.
Since the cascade algorithm approach for the two-direction case
is not much different from that for the one-direction case,
we do not pursue the cascade algorithm approach for $\boldphi$ and $\boldpsi$
in this paper. We pursue only the eigenvalue approach for $\boldphi$
and $\boldpsi$ in this paper.

All calculations of norms, point values, and moments only give the answer
up to an arbitrary constant. This reflects the fact that the refinement equation
only defines $\boldphi$ up to an arbitrary factor.
When we calculate several quantities, how do we make consistent choices?
This is called {\em normalization} and is another issue for the two-direction
multiwavelets which is pursued in this paper.
Normalization of point values and those of derivatives of the one-direction
multiscaling functions and multiwavelets associated with multiscaling functions
can be achieved by fixing the value of some quantity related to multiscaling functions.
We investigate how to normalize point values and those of derivatives
of the two-direction multiscaling functions $\boldphi$ and multiwavelets $\boldpsi$
via the normalization of the zeroth continuous moment of $\boldphi$.


The main objective of this paper is to investigate
the following for orthogonal two-direction multiwavelets:
\begin{itemize}
\item compute point values of orthogonal multiscaling function $\boldphi$ via the eigenvalue approach;
\item determine correct normalization for point values of $\boldphi$ in terms of the zeroth continuous moment $\boldm_0 := \int_{-\infty}^\infty \boldphi(x)\,\textrm{d}x$ of $\boldphi$;
\item compute point values of orthogonal multiwavelets $\boldpsi^{(s)}$, $s=1,2,\ldots,d-1$, associated with $\boldphi$;
\item compute point values of the $n$th derivative $D^{n}\boldphi$ 
      via the eigenvalue approach;
\item determine correct normalization for point values of $D^{n}\boldphi$ in terms of the $j$th continuous moments $\boldm_j := \int_{-\infty}^\infty x^j \boldphi(x)\,\textrm{d}x$ for $j=0,1,\ldots,n$;
\item compute point values of $D^{n}\boldpsi^{(s)}$, $s=1,2,\ldots,d-1$.
\end{itemize}

This paper is organized as follows.
Some basic notions and properties of the two-direction multiscaling functions
$\boldphi$ and multiwavelets $\boldpsi$
are introduced in section ~\ref{sec:2direc}.
In section~\ref{pointvalue}, the main results for finding and normalizing
point values of $\boldphi$ and $\boldpsi$
are introduced.
In section~\ref{derivatives}, the main results for finding and normalizing
point values of derivatives of $\boldphi$  and $\boldpsi$
are introduced.
Finally, examples for illustrating the general theory in sections 2, 3, and 4
are given in section~\ref{examples}.

\bigskip
\section{Two-direction multiscaling functions and multiwavelets}\label{sec:2direc}

In this section we review and investigate orthogonal two-direction multiscaling
functions and orthogonal two-direction multiwavelets
(see ~\citep{YangLi2007AMC,YangLi2007SciChina,YangXie2008}).

A {\em two-direction  multiscaling function} of multiplicity $r$ and dilation
factor $d$ is a vector of $r$ real-valued functions
\begin{equation*}
\boldphi(x) = \left[ \phi_{1}(x), \phi_{2}(x), \ldots, \phi_{r}(x)
\right]^T, \quad x \in \bbR,
\end{equation*}
which satisfies a recursion relation
\begin{equation}\label{recrel}
  \boldphi(x) = \sqrt{d} \,\sum_{k\in\Z} \left[ P^{+}_{k} \,\boldphi(dx-k) + P^{-}_{k} \,\boldphi(k-dx) \right]
\end{equation}
and generates a multiresolution approximation of $L^{2}(\bbR)$.
The $P^{+}_{k}$, $P^{-}_{k}$, called the positive- and negative-direction
recursion coefficients for $\boldphi$, respectively, are $r \times r$ matrices.

For standard one-direction (multi)wavelets, the scaling function is a linear combination
of scaled and shifted versions of itself.
For two-direction (multi)wavelets, it is a linear combination of scaled and shifted
versions of itself and of its reverse.

{\em Two-direction multiwavelets} $\boldpsi^{(s)}$, $s=1,2,\ldots,d-1$, associated
with $\boldphi$ satisfy
\begin{equation}\label{recrelpsi}
  \boldpsi^{(s)}(x) = \sqrt{d} \,\sum_{k\in\Z} \left[ Q^{(s)+}_{k} \,\boldphi(dx-k) + Q^{(s)-}_{k} \,\boldphi(k-dx) \right], \quad s=1,2,\ldots,d-1.
\end{equation}
The $Q^{(s)+}_{k}$, $Q^{(s)-}_{k}$ are called the positive- and negative-direction
recursion coefficients for $\boldpsi^{(s)}$.

By taking Fourier transform on both sides of ~\eqref{recrel}, we have
\begin{equation}\label{ft_recrel}
    \widehat{\boldphi}(\xi) = P^{+}(e^{-i\xi/d}) \widehat{\boldphi}(\xi/d) + P^{-}(e^{-i\xi/d}) \overline{\widehat{\boldphi}(\xi/d)},
\end{equation}
\enlargethispage*{.7\baselineskip}
where
\begin{equation}
    P^{+}(z) = \frac{1}{\sqrt{d}} \,\sum_{k\in\Z} P^{+}_k z^k
    \qquad \text{and} \qquad
    P^{-}(z) = \frac{1}{\sqrt{d}} \,\sum_{k\in\Z} P^{-}_k z^k,
\end{equation}
for $|z|=1$ and $z\in\C$, are called positive- and negative-direction mask symbols,
respectively.

We rewrite the two-direction multiscaling function ~\eqref{recrel} as
\begin{equation}\label{recrelneg}
  \boldphi(-x) = \sqrt{d} \,\sum_{k\in\Z} \left[ P^{+}_{k} \,\boldphi(-dx-k) + P^{-}_{k} \,\boldphi(k+dx) \right]
\end{equation}

By taking Fourier transform on both sides of ~\eqref{recrelneg}, we have
\begin{equation}\label{ft_recrelneg}
    \overline{\widehat{\boldphi}(\xi)} = \overline{P^{+}(e^{-i\xi/d})}\, \overline{\widehat{\boldphi}(\xi/d)} + \overline{P^{-}(e^{-i\xi/d})} \, \widehat{\boldphi}(\xi/d).
\end{equation}
From ~\eqref{ft_recrel} and ~\eqref{ft_recrelneg}, we have
\begin{equation}\label{Phifreq}
    \begin{bmatrix}
      \widehat{\boldphi}(\xi) \\ \noalign{\medskip}
      \overline{\widehat{\boldphi}(\xi)} \\
    \end{bmatrix}
    =
    \begin{bmatrix}
      P^{+}(e^{-i\xi/d}) & P^{-}(e^{-i\xi/d})  \\ \noalign{\medskip}
      \overline{P^{-}(e^{-i\xi/d})} & \overline{P^{+}(e^{-i\xi/d})} \\
    \end{bmatrix}
    \begin{bmatrix}
      \widehat{\boldphi}(\xi/d) \\ \noalign{\medskip}
      \overline{\widehat{\boldphi}(\xi/d)} \\
    \end{bmatrix}.
\end{equation}
We see that ~\eqref{recrel} has a solution if and only if ~\eqref{Phifreq}
has a solution.

Let us combine two-direction multiscaling functions $\boldphi(x)$ and $\boldphi(-x)$ to construct
a one-direction multiscaling function $\boldPhi(x)$ so that all the (one-direction) multiwavelet theory
applies to $\boldPhi(x)$.
Let \begin{equation}\label{recrelPhi}
\boldPhi(x) =
\begin{bmatrix}
  \boldphi(x) \\ \noalign{\smallskip}
  \boldphi(-x) \\
\end{bmatrix}
= \sqrt{d} \sum_{k\in\Z}
\begin{bmatrix}
    P^{+}_k & P^{-}_k \\ \noalign{\medskip}
    P^{-}_{-k} & P^{+}_{-k} \\
\end{bmatrix}
\boldPhi(dx-k)
\end{equation}
be the matrix refinable equation of $\boldPhi$.
Equation ~\eqref{recrelPhi} is called the {\em deduced} $d$-scale matrix refinement
equation of the two-direction refinement equation ~\eqref{recrel}.
Then ~\eqref{Phifreq} is the $d$-scale matrix refinement equation in frequency
domain of $\boldPhi$. Its refinement mask is
\begin{equation}\label{refinemask}
\mathbf{P}_{\boldPhi}(z) = \begin{bmatrix}
      P^{+}(z) & P^{-}(z) \\ \noalign{\medskip}
      \overline{P^{-}(z)} & \overline{P^{+}(z)} \\
    \end{bmatrix}.
\end{equation}
If $P^{+}_{k}$ and $P^{-}_{k}$, $k\in\Z$, are real, then
\begin{equation}\label{refinemaskreal}
    \mathbf{P}_{\boldPhi}(z) = \frac{1}{\sqrt{d}} \sum_{k\in\Z}
        \begin{bmatrix}
            P^{+}_k & P^{-}_k \\ \noalign{\medskip}
            P^{-}_{-k} & P^{+}_{-k} \\
        \end{bmatrix} z^k.
\end{equation}
In this paper we only consider real recursion coefficients $P^{+}_{k}$, $P^{-}_{k}$, $Q^{(s)+}_{k}$,
and $Q^{(s)-}_{k}$ in $\R^{r\times r}$ for $s=1,2,\ldots,d-1$ and $k\in\Z$.

{\bf Orthogonality.}
Two-direction multiscaling function $\boldphi$ and multiwavelet $\boldpsi^{(s)}$, $s=1,2,\ldots,d-1$,
associated with $\boldphi$ are {\em orthogonal} if
for all $j,k\in\Z$, and $s,t=1,2,\ldots,d-1$,
\begin{align*}
    \< \boldphi(x-j), \boldphi(x-k) \> &= \delta_{jk} I_{r}, \\
    \< \boldphi(x-j), \boldphi(k-x) \> &= O_r, \\
    \< \boldpsi^{(s)}(x-j), \boldpsi^{(t)}(x-k) \> &= \delta_{st} \delta_{jk} I_{r}, \\
    \< \boldpsi^{(s)}(x-j), \boldpsi^{(t)}(k-x) \> &= O_r, \\
    \< \boldphi(x-j), \boldpsi^{(s)}(x-k) \> &= O_r, \\
    \< \boldphi(x-j), \boldpsi^{(s)}(k-x) \> &= O_r,
\end{align*}
where $I_{r}$ is the $r \times r$ identity matrix and $O_r$ is the $r\times r$
zero matrix. Here the inner product is defined by
\begin{equation*}
  \< \boldphi, \boldpsi^{(s)} \> = \int \boldphi(x) {\boldpsi^{(s)}}^*(x) \;\textrm{d}x,
\end{equation*}
\enlargethispage*{.7\baselineskip}
where * denotes the complex conjugate transpose.
This inner product is an $r \times r$ matrix.

{\bf Condition E.}
A matrix $A$ satisfies {\em Condition E} if it has a simple eigenvalue of 1,
and all other eigenvalues are smaller than 1 in absolute value.



Condition E for $\boldphi$ defined in ~\eqref{recrel} means that the matrix
\begin{equation}\label{condsymbol}
\begin{bmatrix}
        P^{+}(1) & P^{-}(1) \\
        P^{-}(1) & P^{+}(1) \\
    \end{bmatrix}
= \frac{1}{\sqrt{d}} \sum_{k\in\Z}
    \begin{bmatrix}
        P^{+}_k & P^{-}_k \\ \noalign{\medskip}
        P^{-}_{k} & P^{+}_{k} \\
    \end{bmatrix}
\end{equation}
satisfies Condition E (see ~\citep[Theorem 3]{XieYang2006} for $d=2$).

For the scalar case $r=1$, since the eigenvalues of the matrix in ~\eqref{condsymbol} are
$P^{+}(1)+P^{-}(1)$ and $P^{+}(1)-P^{-}(1)$, Condition E for $\boldphi$
is equivalent to
\begin{equation*}
  P^{+}(1) + P^{-}(1) = 1, \qquad |P^{+}(1) - P^{-}(1)| < 1,
\end{equation*}
or
\begin{equation*}
  P^{+}(1) - P^{-}(1) = 1, \qquad |P^{+}(1) + P^{-}(1)| < 1
\end{equation*}
(see ~\citep[Theorem 2]{YangLi2007AMC}).



{\bf Moments.}
It is well-known that continuous moments of the
one-direction orthogonal (or biorthogonal) multiscaling function and
multiwavelets can be computed if the recurrence coefficients of $\boldphi$
and $\boldpsi$ are given (see ~\citep{Keinert2004, Kwon2009}).

An algorithm for computing continuous moments of the two-direction multiscaling
function $\boldphi$ and multiwavelets $\boldpsi$ was derived in ~\citep{Kwon2012}.
Continuous moments of $\boldphi$ and $\boldpsi$ can be computed if
the recurrence coefficients of $\boldphi$ and $\boldpsi$ are given.
Here we review some results for moments derived from ~\citep{Kwon2012}.

Let us define the $j$th {\em discrete moment} of two-direction multiscaling function
$\boldphi$ by an $r\times r$ matrix
\begin{equation}\label{discmoment2}
M_{j} = \frac{1}{\sqrt{d}} \sum_{k\in\Z} k^{j} [ P^{+}_k + P^{-}_k ], \qquad j=0,1,2,\ldots.
\end{equation}

Now we separate the $j$th discrete moment $M_j$ into positive and negative parts.
We define the $j$th {\em positive} and {\em negative discrete moments},
$M^{+}_{j}$ and $M^{-}_{j}$ respectively, of the discrete moment $M_{j}$ of
$\boldphi$ by
\begin{gather}\label{dmp2def}
    M^{+}_{j} = \frac{1}{\sqrt{d}} \sum_k k^j P^{+}_k, \qquad
    M^{-}_{j} = \frac{1}{\sqrt{d}} \sum_k k^j P^{-}_k,
\end{gather}
for $j=0,1,2,\ldots$ so that $M_{j} = M^{+}_{j} + M^{-}_{j}$.

The $j$th {\em continuous moment} $\boldm_{j}$ of two-direction multiscaling
function $\boldphi$ is a vector of size $r$ defined by
\begin{equation}\label{cm}
    \boldm_{j} = \int_{-\infty}^{\infty} x^{j} \boldphi(x)\;\textrm{d}x, \qquad j=0,1,2,\ldots.
\end{equation}

The following lemma was proved in ~\citep{Kwon2012}.
For the completeness of the paper, we provide it here again.

\begin{lemma}\label{lem:phinormal} {\rm (\citep{Kwon2012})} {\rm (Normalization for $\boldm_0$)}
If a two-direction multiscaling function $\boldphi \in L^{1} \cap L^{2}$ is
orthogonal and Condition E for $\boldphi$ is satisfied, then
\begin{equation} \label{normal_m0}
   \boldm_{0}^{*} \,\boldm_{0} = \frac{1}{2}.
\end{equation} 
\end{lemma}

\begin{proof}
By expanding the constant 1 in the two-direction orthogonal multiscaling
function series, we have
\begin{equation}\label{oneexpansion}
\begin{split}
1 &= \sum_{k\in\Z} [ \< 1, \boldphi(x-k) \> \,\boldphi(x-k)
   + \< 1, \boldphi(k-x) \> \,\boldphi(k-x) ] \\
&= \boldm_{0}^{*} \sum_{k\in\Z} [ \boldphi(x-k) + \boldphi(k-x) ].
\end{split}
\end{equation}
By integrating~\eqref{oneexpansion} on $[0,1]$, we have
\begin{equation*}
1 = \int_{0}^{1} 1 \;\textrm{d}x
= \boldm_{0}^{*} \sum_{k\in\Z} \int_{0}^{1} [ \boldphi(x-k) + \boldphi(k-x) ] \;\textrm{d}x
= \boldm_{0}^{*} \int_{-\infty}^{\infty} [ \boldphi(x) + \boldphi(-x) ] \;\textrm{d}x
= 2\boldm_{0}^{*} \boldm_{0}.
\end{equation*}
Hence, we have equation~\eqref{normal_m0}.
\end{proof}

In the orthogonal two-direction scalar case, the normalization is
\begin{equation}\label{scalarm0}
    m_0 = \frac{\sqrt{2}}{2}.
\end{equation}

\begin{theorem} {\rm (\citep{Kwon2012})}
The continuous and discrete moments of two-direction multiscaling function
$\boldphi$ are related by
\begin{equation}\label{eq:mb:mncalc}
    \boldm_{j} = d^{-j} \sum_{\ell=0}^{j} \binomial{j}{\ell}
    \left[ M^{+}_{j-\ell}+(-1)^\ell M^{-}_{j-\ell} \right] \,\boldm_{\ell}, \quad j=0,1,2,\ldots.
\end{equation}
In particular,
\begin{equation*}
    \boldm_{0} = M_{0} \boldm_{0}.
\end{equation*}

Moreover,
\begin{equation}\label{mandM22}
\boldm_{j} = \left( d^{j} I_{r} - \left[ M^{+}_{0} + (-1)^j M^{-}_{0}  \right] \right)^{-1}
\sum_{\ell=0}^{j-1} \binomial{j}{\ell} \left[ M^{+}_{j-\ell} + (-1)^j M^{-}_{j-\ell}  \right] \boldm_{\ell}, \qquad j=1,2,\ldots,
\end{equation}
where $I_{r}$ is the identity matrix of size $r\times r$.

Once $\boldm_{0}$ has been chosen, all other continuous moments are
uniquely defined and can be computed from these relations.
\label{thm:mb:mncalc}
\end{theorem}

{\bf Approximation Order.}
The two-direction multiscaling function approximation to a function $f$ at
resolution $d^{-j}$ is given by the series
\begin{equation}\label{fseries}
  P_{j}f = \sum_{k\in\Z} [ \<f, \boldphi^{+}_{jk} \> \,\boldphi^{+}_{jk}
  + \<f, \boldphi^{-}_{jk} \> \,\boldphi^{-}_{jk} ],
\end{equation}
where
\begin{gather*}
  \boldphi^{+}_{jk}(x) = d^{-j/2}\,\boldphi(d^{j}x - k)
  \qquad  \text{and} \qquad
  \boldphi^{-}_{jk}(x) = d^{-j/2}\,\boldphi(k - d^{j}x).
\end{gather*}
The two-direction multiscaling function $\boldphi$ provides
{\em approximation order $p$} if
\begin{equation*}
    \| f - P_{n}f \| = O(d^{-np}),
\end{equation*}
whenever $f$ has $p$ continuous derivatives.

The two-direction multiscaling function $\boldphi$ has {\em accuracy $p$}
if all polynomials of degree up to $p-1$ can be expressed locally as linear
combinations of integer shifts of $\boldphi(x)$ and $\boldphi(-x)$. That is,
there exist row vectors $(\boldc^{+}_{j,k})^{*}$ and $(\boldc^{-}_{j,k})^{*}$,
$j=0,\ldots{},p-1$, $k \in \Z$, so that
\begin{equation}\label{approxint}
  x^{j} = \sum_{k\in\Z} [(\boldc^{+}_{j,k})^{*} \boldphi(x-k)
  + (\boldc^{-}_{j,k})^{*} \boldphi(k-x)].
\end{equation}
\enlargethispage*{.7\baselineskip}




Let $\boldy^{+}_{\ell}=\boldc^{+}_{\ell,0}$ and $\boldy^{-}_{\ell}=\boldc^{-}_{\ell,0}$.
It is assumed that at least one of $\boldy^{+}_{0}$ and $\boldy^{-}_{0}$
is not the zero vector. This assumption will be clear after the basic regularity
condition in Theorem~\ref{thm:brc}.
The vectors $\boldy^{+}_{\ell}$ and $\boldy^{-}_{\ell}$ are called the
{\em approximation vectors}.

\begin{theorem}\label{thm:cmaovec}
Approximation vectors and continuous moments of $\boldphi$ are related by
\begin{equation}\label{srvec}
\begin{split}
  \boldy^{+}_{j} &= \boldm_{j}, \\
  \boldy^{-}_{j}  &= (-1)^j \boldy^{+}_{j} = (-1)^j \boldm_{j},
\end{split}
\end{equation}
and
\begin{equation}\label{eq:yp+yn}
\begin{split}
(\boldy^{+}_{j} + \boldy^{-}_{j})^{*} &= [1+(-1)^{j}] \boldm_{j}^{*}, \\
(\boldy^{+}_{j} - \boldy^{-}_{j})^{*} &= [1-(-1)^{j}] \boldm_{j}^{*},
\end{split}
\end{equation}
for $j=0,1,2,\ldots$.
\end{theorem}

\begin{proof}
We multiply equation ~\eqref{approxint} by $\boldphi(x)$ and integrate to obtain
\begin{equation}\label{eq:yp}
    (\boldy^{+}_{j})^{*} = (\boldc^{+}_{j,0})^{*} = \< x^j, \boldphi(x) \> = \boldm_{j}^{*}.
\end{equation}
We multiply equation ~\eqref{approxint} by $\boldphi(-x)$ and integrate to obtain
\begin{equation}\label{eq:yn}
    (\boldy^{-}_{j})^{*} = (\boldc^{-}_{j,0})^{*} = \< x^j, \boldphi(-x) \> = (-1)^{j} \boldm_{j}^{*}.
\end{equation}
By adding and subtracting equations in ~\eqref{srvec}, we obtain ~\eqref{eq:yp+yn}.
\end{proof}

According to~\eqref{srvec}, for $j=0,1,2,\ldots$, $\boldy^{-}_{j}$ can be expressed
in terms of $\boldy^{+}_{j}$, that is, $\boldy^{-}_{j}=(-1)^{j} \boldy^{+}_{j}$.
We use only one notation $\boldy_{j}$ instead of $\boldy^{+}_{j}$ by removing positive,
that is, we let $\boldy_{j}:=\boldy^{+}_{j}$ from now on and do not use $\boldy^{-}_{j}$ any more.

\begin{theorem} \label{thm:approxorder}
The coefficients $\boldc^{+}_{j,k}$ and $\boldc^{-}_{j,k}$ in equation
~\eqref{approxint} have the form
\begin{equation}\label{approxordervec}
\begin{split}
  \boldc^{+}_{j,k} &= \sum_{\ell=0}^{j} {j \choose \ell} k^{j-\ell} \boldy_{\ell}, \\
  \boldc^{-}_{j,k} &= \sum_{\ell=0}^{j} {j \choose \ell} k^{j-\ell} (-1)^\ell \boldy_{\ell},
\end{split}
\end{equation}
where
$\boldy_{j}=\boldc^{+}_{j,0}=(-1)^j\boldc^{-}_{j,0}$,
and $\boldy_{0}$ is not the zero vector.
\end{theorem}

Here $\boldy_{0}$ is the same vector as in Theorem~\ref{thm:brc}.

\begin{proof}
Replace $x^j$ by $(x+k)^j$ in equation ~\eqref{approxint} and expand.
\end{proof}

A high approximation order is desirable in applications. A minimum
approximation order of 1 is a required condition in many theorems.

The following theorem will be used in proving theorem ~\ref{thm:normalptphi}.

\begin{theorem} \label{thm:brc} {\rm (\citep{Kwon2011c})}
Assume that two-direction multiscaling function $\boldphi$ is a compactly
supported $L^2$-solution of the refinement equations with nonzero integral
and linearly independent shifts. Then 
there exists a nonzero vector $\boldy_{0}\in \R^r$ such that
\begin{equation}\label{brc(iii)}
   \boldy_{0}^{*} \sum_{k\in\Z} [  \boldphi(x-k) + \boldphi(k-x) ] = c,
\end{equation}
where $c$ is a constant.

Moreover, the nonzero vector $\boldy_{0}$ is related to
the zeroth continuous moment of $\boldphi$ by
\begin{equation*}
\boldy_{0}=\boldm_{0}.
\end{equation*}
\end{theorem}

\begin{proof}
Sine the two-direction recursion relation ~\eqref{recrel} has a compactly supported,
$L^2(\R)$-stable solution vector $\boldphi$ of size $r$, its deduced $d$-scale matrix
refinement equation ~\eqref{recrelPhi} has a one-direction compactly supported,
$L^2(\R)$-stable solution vector $\boldPhi$ of size $2r$.

It is well-known in multiwavelet theory that $\boldPhi$ satisfies the following
condition (see~\citep{Keinert2004,Plonka1998}):
There exists a nonzero vector $\boldy\in \R^{2r}$ such that
\begin{equation*}
    \boldy^{*} \sum_{k\in\Z} \boldPhi(x-k) = c,
\end{equation*}
where $c$ is a constant.


Let $\boldy_{0}\in\R^{r}$ be a vector of the upper-half part of
$\boldy\in \R^{2r}$ in (b). By ~\eqref{eq:yp} and ~\eqref{eq:yn}, we have
a structure of $\boldy$ such that
\begin{equation}\label{y_structure}
\boldy^{*} = \< 1, \boldPhi \> = [ \boldm_{0}^{*},  \boldm_{0}^{*}] = [\boldy^{*}_{0}, \boldy^{*}_{0}].
\end{equation}
Since $\boldy$ is a nonzero vector, $\boldy_{0}$ is also a nonzero vector
by ~\eqref{y_structure}.

By ~\eqref{y_structure} and \eqref{brc(iii)}, we have
\begin{equation*}
     \boldy_{0}^{*} \sum_{k\in\Z} [ \boldphi(x-k) + \boldphi(k-x) ]
     = [\boldy_{0}^{*}, \boldy_{0}^{*}] \sum_{k\in\Z}  \begin{bmatrix}
                     \boldphi(x-k) \\ \boldphi(k-x) \\
                   \end{bmatrix}
      = \boldy^{*} \sum_{k\in\Z} \boldPhi(x-k) = c,
\end{equation*}
where $c$ is a constant. This completes the proof.
\end{proof}





Throughout this paper we assume that the two-direction multiscaling function
$\boldphi$ is orthogonal, has compact support, is continuous
(which implies approximation order at least 1), and satisfies
Condition E.

\bigskip
\section{Point values and normalization of the two-direction multiwavelets}
\label{pointvalue}

In multiwavelet theory, computation of point values and those of derivatives for
the one- or two-direction multiscaling functions and multiwavelets is important,
since it provides a method for plotting multiscaling functions and multiwavelets.
It is well-known that point values and normalization of the one-direction
multiscaling function and multiwavelets can be computed if the recurrence
coefficients of multiscaling function and multiwavelets are given
(see ~\citep{Keinert2004}).

In this section we investigate how to compute point values and their normalization
of the orthogonal two-direction multiscaling functions $\boldphi$ and
multiwavelets $\boldpsi^{(s)}$, $s=1,2,\ldots,d-1$, in terms of the recurrence
coefficients $P^{+}_k$, $P^{-}_k$ of $\boldphi$ and $Q^{(s)+}_k$, $Q^{(s)-}_k$
of $\boldpsi^{(s)}$.


The {\em cascade algorithm approach} is a practical way for finding approximate
point values of $\boldphi$ by iteration: for $n=0,1,2,\ldots$,
\begin{equation}\label{cascade}
\boldphi^{(n+1)}(x) = \sqrt{d} \,\sum_{k\in\Z}
\left[ P^{+}_{k} \,\boldphi^{(n)}(dx-k) + P^{-}_{k} \,\boldphi^{(n)}(k-dx) \right],
\end{equation}
where $\boldphi^{(0)}(x) = \max\{1-|x|, 0\}$ is the hat function and $\boldphi^{(n)}$
stands for $\boldphi$ at the $n$th iteration.

There is another approach, called the {\em eigenvalue approach}, which
produces exact point values of $\boldphi$ and $\boldpsi^{(s)}$, $s=1,2,\ldots,d-1$.
It usually works, but may fail in some cases.

Since the cascade algorithm approach for the two-direction case is not much different
from that for the one-direction case, we do not pursue the cascade algorithm approach
but the eigenvalue approach in this section.

In this section we investigate how to
\begin{itemize}
\item compute point values of orthogonal two-direction multiscaling function $\boldphi$ via the eigenvalue approach;
\item normalize point values of $\boldphi$;
\item compute point values of $\boldpsi^{(s)}$, $s=1,2,\ldots,d-1$.
\end{itemize}
\enlargethispage*{.7\baselineskip}


\medskip
{\bf Point values of the two-direction multiscaling functions}

Let $N$ be the largest positive integer such that 
$P^{+}_{-N}$, $P^{+}_{N}$, $P^{-}_{-N}$, or $P^{-}_{N}$ in ~\eqref{recrel}
are nonzero matrices of size $r\times r$.
Then $\boldphi$ is a two-direction multiscaling function satisfying
the refinement equation
\begin{equation}\label{recrelcpt}
  \boldphi(x) = \sqrt{d} \, \sum_{k=-N}^{N}  \left[ P^{+}_{k} \,\boldphi(dx-k) + P^{-}_{k} \,\boldphi(k-dx) \right].
\end{equation}
Suppose that $\boldphi$ in ~\eqref{recrelcpt} satisfies Condition E for $\boldphi$.
Then it is easy to prove that the support of $\boldphi$ in ~\eqref{recrelcpt}
is also contained in
$[-N/(d-1), N/(d-1)]$. (Proof is given for
\begin{equation*}
  \boldphi(x) = \sqrt{d} \, \left[ \sum_{k=0}^{N}  P^{+}_{k} \,\boldphi(dx-k) + \sum_{k=-N}^{0} P^{-}_{k} \,\boldphi(k-dx) \right]
\end{equation*}
in~\citep{YangLi2007SciChina}, Theorem 4).

Let $a$ and $b$ be the smallest and largest integers in this interval.

For integer $\ell$, $a \le \ell \le b$, the refinement
equation~\eqref{recrelcpt} leads to
\begin{equation*}
\begin{aligned}
  \boldphi(\ell) &= \sqrt{d} \, \sum_{k=-N}^{N} \left[  P^{+}_{k} \,\boldphi(d\ell-k) + P^{-}_{k} \,\boldphi(k-d\ell) \right] \\
  &= \sqrt{d} \left[ \sum_{k=d\ell-N}^{d\ell} P^{+}_{d\ell-k} \boldphi(k) + \sum_{k=-d\ell}^{N-d\ell} P^{-}_{d\ell+k} \boldphi(k) \right].
\end{aligned}
\end{equation*}
Since $\boldphi(k)=0$ for every $k \in \Z\setminus[a,b]$, we have
\begin{equation}\label{recrelcpt2}
  \boldphi(\ell) = \sqrt{d} \sum_{k=a}^{b} \left[ P^{+}_{d\ell-k} + P^{-}_{d\ell+k} \right]\, \boldphi(k), \qquad a\leq \ell \leq b.
\end{equation}
This is an eigenvalue problem
\begin{equation}\label{eigencpt}
  \begin{bmatrix}
    \boldphi(a) \\ \boldphi(a+1) \\ \vdots \\ \boldphi(b)
  \end{bmatrix}
  = T_\boldphi \begin{bmatrix}
    \boldphi(a) \\ \boldphi(a+1) \\ \vdots \\ \boldphi(b)
  \end{bmatrix},
\end{equation}
where
\begin{equation}\label{T_phi}
  (T_{\boldphi})_{\ell k} = \sqrt{d} \,(P^{+}_{d\ell-k} + P^{-}_{d\ell+k}),
  \quad a \le \ell, k \le b.
\end{equation}
We note that each column of $T_\boldphi$ contains all of the
$P^{+}_{dk-\ell} + P^{-}_{dk+\ell}$ for some fixed $\ell$.
The basic regularity condition (iii) in Theorem~\ref{thm:brc} implies that
$(\boldy_{0}^{*},\boldy_{0}^{*},\ldots{},\boldy_{0}^{*})$ 
is a left eigenvector to eigenvalue 1, so a right eigenvector also exists.

We assume that this eigenvalue is simple, so that the solution is
unique. (This is the place where the algorithm could fail).

Let $k_1$ and $k_2$ be the smallest and largest integers such that $P^{+}_{k_1}$ and $P^{+}_{k_2}$ are nonzero matrices, respectively.
Let $k_3$ and $k_4$ be the smallest and largest integers such that $P^{-}_{k_3}$ and $P^{-}_{k_4}$ are nonzero matrices, respectively.
In the case of dilation $d=2$, the support of $\boldphi$ is contained in $[a,b]\subset[-N,N]$.
The first row of matrix equation~\eqref{eigencpt} is either
\begin{align*}
  \boldphi(a) &= \sqrt{d} \, P^{+}_{k_1} \boldphi(a) \qquad \text{or} \qquad
  \boldphi(a) = \sqrt{d} \, P^{-}_{k_3} \boldphi(b). 
\end{align*}
The last row of matrix equation~\eqref{eigencpt} is either
\begin{align*}
  \boldphi(b) &= \sqrt{d} \, P^{+}_{k_2} \boldphi(b) \qquad \text{or} \qquad
  \boldphi(b) = \sqrt{d} \, P^{-}_{k_4} \boldphi(a). 
\end{align*}

For the case that the first row is $\boldphi(a) = \sqrt{d} \, P^{+}_{k_1} \boldphi(a)$, unless $P^{+}_{k_1}$ 
has an eigenvalue of $1/\sqrt{d}$, the value of $\boldphi$ at the left endpoint
is zero, and we can reduce the size of
$
\begin{bmatrix}
    \boldphi(a) & \boldphi(a+1) & \hdots & \boldphi(b)
\end{bmatrix}^T
$
and $T_\boldphi$.
For the case that the last row is $\boldphi(b) = \sqrt{d} \, P^{+}_{k_2} \boldphi(b)$, unless $P^{+}_{k_2}$ 
has an eigenvalue of $1/\sqrt{d}$, the value of $\boldphi$ at the right endpoint
is zero, and we can reduce the size of
$ 
\begin{bmatrix}
    \boldphi(a) & \boldphi(a+1) & \hdots & \boldphi(b)
\end{bmatrix}^T
$ 
and $T_\boldphi$.

All calculations of norms, point values, and moments only give the answer
up to an arbitrary constant. This reflects the fact that the
refinement equation only defines $\boldphi$ up to an arbitrary
factor. When we calculate several quantities, how do we make
consistent choices? This is called \emph{normalization}
and an issue to be pursued in this section.

\medskip
{\bf Normalization of point values of the two-direction multiscaling function}

The normalization for point values of $\boldphi$ is given in the following Theorem. 

\begin{theorem}\label{thm:normalptphi}
{\rm (Normalization of point values of $\boldphi$).}
The correct normalization for point values of two-direction multiscaling function
$\boldphi$ is
\begin{equation}\label{normalptphi2}
    \boldm_{0}^{*}\left( \sum_{k\in\Z} \boldphi(k) \right)
= \boldm_{0}^{*} \boldm_{0}.
\end{equation}
\end{theorem}

\begin{proof}
If $\boldphi$  satisfies the basic regularity conditions, then we have, by setting
$x=0$ in ~\eqref{brc(iii)}, 
\begin{equation*}
c =  2(\boldy_0)^* \left( \sum_{k\in\Z} \boldphi(k) \right)
= 2(\boldm_0)^* \left( \sum_{k\in\Z} \boldphi(k) \right).
\end{equation*}
By integrating equation ~\eqref{brc(iii)} in theorem ~\ref{thm:brc} on $[0,1]$,
we have
\begin{equation*}
c = \int_{0}^{1} c \;\textrm{d}x
= \boldy_{0}^{*} \sum_{k\in\Z} \int_{0}^{1} [ \boldphi(x-k) + \boldphi(k-x) ] \;\textrm{d}x
= \boldy_{0}^{*} \int_{-\infty}^{\infty} [ \boldphi(x) + \boldphi(-x) ] \;\textrm{d}x
= 2\boldm_{0}^{*} \boldm_{0}.
\end{equation*}

Hence, 
we have ~\eqref{normalptphi2}.
\end{proof}

According to equation~\eqref{normalptphi2}, unlike the scalar case, the sum of
point values at the integers and the integral do not have to be the same.
They just have to have the same inner product with $\boldm_{0}$.

For the orthogonal two-direction scalar case, the correct normalization for point
values of $\phi$ is
\begin{equation}\label{normalptphir1}
    \sum_{k\in\Z} \phi(k) = \frac{\sqrt{2}}{2}.
\end{equation}

Once the values of $\boldphi$ at the integers have been determined, we can use
the refinement equation~\eqref{recrelcpt} to obtain values of $\boldphi$ at points
of the form $k/d$, $k\in\Z$, then $k/d^{2}$, and so on to any desired resolution.

\medskip
{\bf Point values of the two-direction multiwavelets}


Suppose that $\boldpsi^{(s)}$, $s=1,2,\ldots,d-1$, are the two-direction multiwavelets
associated with $\boldphi$ defined by
\begin{equation}\label{recrelpsicpt}
\begin{aligned}
  \boldpsi^{(s)}(x) &= \sqrt{d} \sum_{k=-N}^{N} \left[ Q^{(s)+}_{k} \, \boldphi(dx-k) + Q^{(s)-}_{k} \, \boldphi(k-dx) \right]
\end{aligned}
\end{equation}
for some positive integer $N$.

For $s=1,2,\ldots,d-1$ and integer $\ell$, $a \le \ell \le b$,
the refinement equation~\eqref{recrelpsicpt} leads to
\begin{equation*}
\begin{aligned}
  \boldpsi^{(s)}(\ell) &= \sqrt{d} \sum_{k=-N}^{N} \left[ Q^{(s)+}_{k} \, \boldphi(d\ell-k) + Q^{(s)-}_{k} \, \boldphi(k-d\ell) \right] \\
  &= \sqrt{d} \left[ \sum_{k=d\ell-N}^{d\ell} Q^{(s)+}_{d\ell-k} \boldpsi(k) + \sum_{k=-d\ell}^{N-d\ell} Q^{(s)-}_{k-d\ell} \boldpsi(k) \right].
\end{aligned}
\end{equation*}
Since $\boldphi(k) =0$ for every $k \in \Z\setminus[a,b]$, we have
\begin{equation}\label{recrelpsicpt2}
  \boldpsi^{(s)}(\ell) = \sqrt{d} \sum_{k=a}^{b} \left[ Q^{(s)+}_{d\ell-k} + Q^{(s)-}_{k-d\ell} \right]\, \boldphi(k), \qquad a\leq \ell \leq b.
\end{equation}
This is a multiplication
\begin{equation}\label{multpsi}
  \begin{bmatrix}
    \boldpsi^{(s)}(a) \\ \boldpsi^{(s)}(a+1) \\ \vdots \\ \boldpsi^{(s)}(b)
  \end{bmatrix}
  = T_{\boldpsi^{(s)}} \begin{bmatrix}
    \boldphi(a) \\ \boldphi(a+1) \\ \vdots \\ \boldphi(b)
  \end{bmatrix},
\end{equation}
where
\begin{equation}\label{T_psi}
  (T_{\boldpsi^{(s)}})_{\ell k}  = \sqrt{d} \,(Q^{(s)+}_{d\ell-k} + Q^{(s)-}_{d\ell+k}),
  \quad a \le \ell, k \le b.
\end{equation}
Note that each column of $T_{\boldpsi^{(s)}}$ contains all of the
$Q^{(s)+}_{dk-\ell} + Q^{(s)-}_{dk+\ell}$ for some fixed $\ell$.

Once the values of $\boldphi$ at the integers have been determined, we can use
the refinement equation~\eqref{recrelpsicpt} to obtain values of $\boldpsi^{(s)}$
at points of the form $k/d$, $k\in\Z$, then $k/d^{2}$, and so on to any desired resolution.


\bigskip
\section{Point values and normalization of derivatives of the two-direction multiwavelets} \label{derivatives}

In this section we investigate how to compute point values of derivatives
of two-direction multiscaling function $\boldphi$ and two-direction multiwavelets
$\boldpsi^{(s)}$, $s=1,2,\ldots,d-1$ associated with $\boldphi$.

The main idea of the method is similar to the eigenvalue problem in Section 3.

We can  use the eigenvalue approach to compute point values of derivatives
of $\boldphi$ (assuming they exist).

In this section we investigate how to
\begin{itemize}
\item compute point values of the $n$th derivative $D^{n}\boldphi$ 
      via the eigenvalue approach;
\item determine correct normalization for point values of $D^{n}\boldphi$ in terms of the $j$th continuous moments $\boldm_j := \int_{-\infty}^\infty x^j \boldphi(x)\,\textrm{d}x$ for $j=0,1,\ldots,n$;
\item compute point values of $D^{n}\boldpsi^{(s)}$, $s=1,2,\ldots,d-1$.
\end{itemize}

\medskip
{\bf Point values of derivatives of the two-direction multiscaling functions}

Let us recall ~\eqref{recrelcpt} that $\boldphi$ satisfies the refinement equation
\begin{equation*}
  \boldphi(x) = \sqrt{d} \,\sum_{k=-N}^{N} \left[ P^{+}_{k} \,\boldphi(dx-k) + P^{-}_{k} \,\boldphi(k-dx) \right]
\end{equation*}
for some positive integer $N$. Then
\begin{equation}\label{recrelDncpt}
  (D^{n}\boldphi)(x) = d^{n} \sqrt{d} \,\sum_{k=-N}^{N} \left[ P^{+}_{k} \,(D^{n}\boldphi)(dx-k) + (-1)^n P^{-}_{k} \,(D^{n}\boldphi)(k-dx) \right].
\end{equation}
For integer $\ell$, $a \le \ell \le b$, the refinement equation~\eqref{recrelDncpt}
leads to
\begin{equation*}
\begin{aligned}
  (D^{n}\boldphi)(\ell) &= d^{n} \sqrt{d} \,\sum_{k=-N}^{N} \left[ P^{+}_{k} \,(D^{n}\boldphi)(d\ell-k) + (-1)^n P^{-}_{k} \,(D^{n}\boldphi)(k-d\ell) \right] \\
  &= d^{n} \sqrt{d} \left[ \sum_{k=d\ell-N}^{d\ell} P^{+}_{d\ell-k} D^{n}\boldphi(k) + (-1)^n \sum_{k=-d\ell}^{N-d\ell} P^{-}_{d\ell+k} D^{n}\boldphi(k)\right].
\end{aligned}
\end{equation*}
Since $D^{n}\boldphi(k)=0$ for every $k \in \Z\setminus[a,b]$, we have
\begin{equation}\label{recrelDncpt2}
  (D^{n}\boldphi)(\ell) = d^{n} \sqrt{d} \sum_{k=a}^{b} \left[ P^{+}_{d\ell-k} + (-1)^n P^{-}_{d\ell+k} \right]\, D^{n}\boldphi(k).
\end{equation}

This is an eigenvalue problem
\begin{equation}\label{eigenphiDn}
    D^{n}\boldphi = d^{n} \,T_{D^{n}\boldphi} \,D^{n}\boldphi,
\end{equation}
where
\begin{equation}\label{T_Dphi}
  (T_{D^{n}\boldphi})_{\ell k} = \sqrt{d} \,(P^{+}_{d\ell-k} + (-1)^n P^{-}_{d\ell+k}),
  \quad a \le \ell, k \le b.
\end{equation}
So $D^{n}\boldphi$ is an eigenvector of $T_{D^{n}\boldphi}$ to the eigenvalue $1/d^{n}$.
This eigenvalue must exist if $\boldphi$ is $n$ times differentiable.

\medskip
{\bf Normalization of point values of the $n$th derivatives of the two-direction
multiscaling function}

The normalization for point values of the $n$th derivative $D^{n}\boldphi$ is given
in the following Theorem.

\begin{theorem} \label{thm:derivnormphi1}
{\rm (Normalization for point values of the $n$th derivative $D^{n}\boldphi$).}
The correct normalization for the $n$th derivative $D^n \boldphi$ of the
two-direction multiscaling function $\boldphi$ is given by
\begin{equation}\label{eq:derivnormphi1}
\frac{n!}{2} = \sum_{\ell=0
}^{n} \binomial{n}{\ell} (-1)^{n-\ell}
\boldm_{\ell}^{*} \left( \sum_{k\in\Z} k^{n-\ell} D^{n}\boldphi(k) \right).
\end{equation}
\end{theorem}

\begin{proof}
It is well-known in multiwavelet theory that if $\boldphi$ has $n$ continuous derivatives, it has approximation order at least $n+1$.
By ~\eqref{approxint} and Theorem~\ref{thm:approxorder}, we have 
\begin{align*}
x^{n} &= \sum_{k\in\Z} [ (\boldc^{+}_{n,k})^{*} \boldphi(x-k) + (\boldc^{-}_{n,k})^{*} \boldphi(k-x)] \\
&= \sum_{\ell=0}^{n} \sum_{k\in\Z} \binomial{n}{\ell} k^{n-\ell}
(\boldy_{\ell})^{*} \left[ \boldphi(x-k)
+ (-1)^\ell \boldphi(k-x) \right].
\end{align*}
By differentiating $n$ times and setting $x=0$ on both sides of the above equation,
we have
\begin{align*}
n! &= \sum_{\ell=0}^{n} \sum_{k\in\Z} \binomial{n}{\ell} k^{n-\ell}
(\boldy_{\ell})^{*} \left[ D^{n}\boldphi(-k)
+ (-1)^{n+\ell} D^{n}\boldphi(k) \right] \\
&= \sum_{\ell=0}^{n} \binomial{n}{\ell} (-1)^{n-\ell} (\boldy_{\ell})^{*}
[ 1 + (-1)^{2\ell} ] \left( \sum_{k\in\Z} k^{n-\ell} D^{n}\boldphi(k) \right) \\
&= 2 \sum_{\ell=0}^{n} 
\binomial{n}{\ell} (-1)^{n-\ell}
\boldm_{\ell}^{*} \left( \sum_{k\in\Z} k^{n-\ell} D^{n}\boldphi(k) \right).
\end{align*}
\end{proof}

For future reference, we list some in detail. Normalization for the $n$th
derivative $D^{n}\boldphi$ of $\boldphi$ for $n=1,2,3$ are
\begin{eqnarray*}
\frac{1}{2} &=& - \boldm_0^* \left( \sum_{k\in\Z} k D\boldphi(k) \right)
+ \boldm_1^* \left( \sum_{k\in\Z} D\boldphi(k) \right),\\
1 &=& \boldm_0^*
    \left( \sum_{k\in\Z} k^2 D^{2}\boldphi(k) \right)
    - 2\boldm_1^* \left( \sum_{k\in\Z} k D^{2}\boldphi(k) \right) + \boldm_2^*
    \left( \sum_{k\in\Z} D^{2}\boldphi(k) \right), \\
3 &=& - \boldm_0^* \left( \sum_{k\in\Z} k^3 D^{3}\boldphi(k) \right)
    + 3\boldm_1^* \left( \sum_{k\in\Z} k^2 D^{3}\boldphi(k) \right)
    - 3\boldm_2^* \left( \sum_{k\in\Z} k D^{3}\boldphi(k) \right) \\
    & & + \boldm_3^* \left( \sum_{k\in\Z} D^{3}\boldphi(k) \right).
\end{eqnarray*}

For the orthogonal two-direction scalar case, the correct normalization for
point values of $D\phi$ is
\begin{equation} \label{eq:derivnormphi1scalar}
\frac{1}{2} = - \frac{\sqrt{2}}{2} \left( \sum_{k\in\Z} k D\phi(k) \right)
+ m_1^* \left( \sum_{k\in\Z} D\phi(k) \right).
\end{equation}

Once the values of $D^{n}\boldphi$ at the integers have been determined, we can use
the refinement equation~\eqref{recrelDncpt} to obtain values of $D^{n}\boldphi$
at points of the form $k/d$, $k \in \Z$, then $k/d^{2}$,
and so on to any desired resolution.

\medskip
{\bf Point values of derivatives of the two-direction multiwavelets}

Now we work on point values of derivatives for two-direction multiwavelets.

Let us recall ~\eqref{recrelpsicpt} that $\boldpsi^{(s)}$, $s=1,2,\ldots,d-1$,
satisfy the refinement equation
\begin{equation*}
  \boldpsi^{(s)}(x) = \sqrt{d} \,\sum_{k=-N}^{N} \left[ Q^{(s)+}_{k} \,\boldphi(dx-k) + Q^{(s)-}_{k} \,\boldphi(k-dx) \right]
\end{equation*}
for some positive integer $N$. Then
\begin{equation}\label{recrelpsiDncpt}
  (D^{n}\boldpsi^{(s)})(x) = d^n \sqrt{d} \,\sum_{k=-N}^{N} \left[ Q^{(s)+}_{k} \,(D^{n}\boldphi)(dx-k) + (-1)^n Q^{(s)-}_{k} \,(D^{n}\boldphi)(k-dx) \right].
\end{equation}
For $s=1,2,\ldots,d-1$ and integer $\ell$, $a \le \ell \le b$, the refinement equation~\eqref{recrelpsiDncpt}
leads to
\begin{equation*}
\begin{aligned}
  (D^{n}\boldpsi^{(s)})(\ell) &= d^n \sqrt{d} \,\sum_{k=-N}^{N} \left[ Q^{(s)+}_{k} \,(D^{n}\boldphi)(dx-k) + (-1)^n Q^{(s)-}_{k} \,(D^{n}\boldphi)(k-dx) \right] \\
  &= d^n \sqrt{d} \left[ \sum_{k=d\ell-N}^{d\ell} Q^{(s)+}_{d\ell-k}  (D^{n}\boldphi)(k)
  + (-1)^n \sum_{k=-d\ell}^{N-d\ell} Q^{(s)-}_{d\ell+k}  (D^{n}\boldphi)(k) \right].
\end{aligned}
\end{equation*}
Since $D^{n}\boldpsi^{(s)}(k)=0$ for every $k \in \Z\setminus[a,b]$, we have
\begin{equation}\label{recrelpsiDncpt2}
  (D^{n}\boldpsi^{(s)})(\ell) = d^n \sqrt{d} \sum_{k=a}^{b} \left[ Q^{(s)+}_{d\ell-k} + (-1)^n Q^{(s)-}_{d\ell+k} \right]\, (D^{n}\boldphi)(k).
\end{equation}

This is a multiplication
\begin{equation}\label{multpsiDn}
    D^{n}\boldpsi^{(s)} = d^n \,T_{D^{n}\psi^{(s)}} \,D^{n}\boldphi,
\end{equation}
where
\begin{equation}\label{T_Dnpsi}
  (T_{D^{n}\psi^{(s)}})_{\ell k}
  = \sqrt{d} \,(Q^{(s)+}_{d\ell-k} + (-1)^n Q^{(s)-}_{d\ell+k}), \quad a \le \ell, k \le b.
\end{equation}
This $D^{n}\boldpsi^{(s)}$ must exist if $D^{n}\boldphi$ exists.


Once the values of $D^{n}\boldphi$ at the integers have been determined, we can use
the refinement equation~\eqref{recrelpsiDncpt} to obtain values of $D^{n}\psi^{(s)}$
at points of the form $k/d$, $k \in \Z$, then $k/d^{2}$,
and so on to any desired resolution.

%

\bigskip
\section{Examples}\label{examples}

In this section, we give two examples for illustrating the general theory
in sections 2, 3, and 4.

\bigskip
\begin{example} \label{ex1} {\rm
In this example we take orthogonal two-direction scaling function $\phi$ of
approximation order 2
derived from BAT O2 (orthogonal balanced one-direction multiscaling function of order 2
~\citep{Lebrun-Vetterli-1998b,Lebrun-Vetterli-2001}) given in ~\citep{Kwon2011d}.
We note that multiplicity $r=1$ and dilation factor $d=2$.

The nonzero recursion coefficients for $\phi$ are
\begin{gather*}
    P^{+}_{1} = \frac{93-13\sqrt{31}}{640\sqrt{2}}, \quad
    P^{+}_{2} = \frac{341-11\sqrt{31}}{640\sqrt{2}}, \quad
    P^{+}_{3} = \frac{11-11\sqrt{31}}{640\sqrt{2}}, \quad
    P^{+}_{4} = \frac{-13+3\sqrt{31}}{640\sqrt{2}}, \\
    P^{-}_{4} = \frac{-31+\sqrt{31}}{640\sqrt{2}}, \quad
    P^{-}_{5} = \frac{217+23\sqrt{31}}{640\sqrt{2}}, \quad
    P^{-}_{6} = \frac{23+7\sqrt{31}}{640\sqrt{2}}, \quad
    P^{-}_{7} = \frac{-1+\sqrt{31}}{640\sqrt{2}}. 
\end{gather*}

The nonzero recursion coefficients for $\psi$ are
\begin{gather*}
    Q^{+}_{1} = \frac{11-\sqrt{31}}{160\sqrt{2}}, \quad
    Q^{+}_{2} = \frac{57+3\sqrt{31}}{160\sqrt{2}}, \quad
    Q^{+}_{3} = \frac{-91+\sqrt{31}}{160\sqrt{2}}, \quad
    Q^{+}_{4} = \frac{23-3\sqrt{31}}{160\sqrt{2}}, \\
    Q^{-}_{4} = Q^{+}_{4}, \quad
    Q^{-}_{5} = Q^{+}_{3}, \quad
    Q^{-}_{6} = Q^{+}_{2}, \quad
    Q^{-}_{7} = Q^{+}_{1}, 
\end{gather*}

Condition E for $\phi$ is satisfied.
$\phi$ and $\psi$ are supported on $[1/2,7/2]\subset [0,4]$.
$\phi$ is orthogonal and $\psi$ is also orthogonal.

Sobolev exponent of $\phi$ is 1.6310, same as that for BAT O2 (orthogonal balanced one-direction multiscaling function of order 2) given in ~\citep{Lebrun-Vetterli-1998b,Lebrun-Vetterli-2001}.

By ~\eqref{scalarm0} and ~\eqref{mandM22}
the $j$th continuous moments $m_j$ of $\phi$ for $j=0,1,2$ are
\begin{equation*}
m_0 = \frac{\sqrt{2}}{2} \approx 0.7071,\qquad
m_1 = \frac{7\sqrt{2}}{8} \approx 1.2374 \qquad
m_2 = \frac{49\sqrt{2}}{32} \approx 2.1655.
\end{equation*}

\medskip
{\bf Point values of $\phi$}\\
The eigenvalue problem in~\eqref{eigencpt} is
\begin{equation*}
  \begin{bmatrix}
    \phi(0) \\ \phi(1) \\ \phi(2) \\ \phi(3) \\ \phi(4)
  \end{bmatrix}
  = T_\phi \begin{bmatrix}
    \phi(0) \\ \phi(1) \\ \phi(2) \\ \phi(3) \\ \phi(4)
  \end{bmatrix},
\end{equation*}
where
\begin{equation*}
T_\phi = T_{\phi^{+}} + T_{\phi^{-}}
= \begin{bmatrix}
  0 & 0 & 0 & 0 & P^{-}_{4} \\
  P^{+}_{2} & P^{+}_{1} & P^{-}_{4} & P^{-}_{5}  & P^{-}_{6} \\
  P^{+}_{4}+P^{-}_{4} & P^{+}_{3}+P^{-}_{5} & P^{+}_{2}+P^{-}_{6} & P^{+}_{1}+P^{-}_{7} & 0  \\
  P^{-}_{6} & P^{-}_{7} & P^{+}_{4} & P^{+}_{3} & P^{+}_{2} \\
  0 &  0 & 0 & 0 & P^{+}_{4}  \\
\end{bmatrix}.
\end{equation*}
The matrix $T_\phi$ has the eigenvalues $1, -0.1783, 0.1536, 0.0116$, and $0$.
The matrix $T_\phi$ has the eigenvector
\begin{equation*}
\begin{bmatrix}
    \phi(0) & \phi(1) & \phi(2) & \phi(3) & \phi(4)
\end{bmatrix}^T
\approx \begin{bmatrix}
     0 &  -0.0564 & 0.7566 & 0.0069 & 0
\end{bmatrix}^T
\end{equation*}
to the eigenvalue 1.

Since the correct normalization for $\phi$ is
\begin{equation*}
    \sum_{k\in\Z} \phi(k) = \frac{\sqrt{2}}{2},
\end{equation*}
the normalizing constant is $1$.
Hence, $\phi$ is normalized already.
Point values of normalized $\phi$ at integers are
\begin{equation*}
\begin{bmatrix}
    \phi(0) & \phi(1) & \phi(2) & \phi(3) & \phi(4)
\end{bmatrix}^T
\approx \begin{bmatrix}
     0 &  -0.0564 & 0.7566 & 0.0069 & 0
\end{bmatrix}^T.
\end{equation*}

Once the values of $\phi$ at the integers have been determined, we can use
the refinement equation~\eqref{recrelcpt} to obtain values at points of the form
$k/d$, $k \in \Z$, then $k/d^{2}$, and so on to any desired resolution.


\medskip
{\bf Point values of $\psi$}\\
The multiplication in~\eqref{multpsi} is
\begin{equation*}
  \begin{bmatrix}
    \psi(0) \\ \psi(1) \\ \psi(2) \\ \psi(3) \\ \psi(4)
  \end{bmatrix}
  = T_\psi \begin{bmatrix}
    \phi(0) \\ \phi(1) \\ \phi(2) \\ \phi(3) \\ \phi(4)
  \end{bmatrix},
\end{equation*}
where
\begin{equation*}
T_\boldpsi = \begin{bmatrix}
  0 & 0 & 0 & 0 & Q^{-}_{4} \\
  Q^{+}_{2} & Q^{+}_{1} & Q^{-}_{4} & Q^{-}_{5} & Q^{-}_{6} \\
  Q^{+}_{4}+Q^{-}_{4} & Q^{+}_{3}+Q^{-}_{5} & Q^{+}_{2}+Q^{-}_{6} & Q^{+}_{1}+Q^{-}_{7} & 0 \\
  Q^{-}_{6} & Q^{-}_{7} & Q^{+}_{4} & Q^{+}_{3} & Q^{+}_{2} \\
  0 &  0 & 0 & 0 & Q^{+}_{4}  \\
\end{bmatrix}.
\end{equation*}
Hence, we have
\begin{equation*}
\begin{bmatrix}
   \psi(0) & \psi(1) & \psi(2) & \psi(3) & \psi(4)
\end{bmatrix}^T
\approx \begin{bmatrix}
    0 & 0.0484 & 1.5154 & 0.0484 & 0
\end{bmatrix}^T.
\end{equation*}

Once the values of $\phi$ and $\boldpsi$ at the integers have been determined,
we can use the equation~\eqref{recrelpsicpt2} to obtain values of
$\boldpsi$ at points of the form $k/d$, $k \in \Z$, then $k/d^{2}$, and so on
to any desired resolution.


See Figure ~\ref{fig:bat2to2dir} for the graphs of $\phi$ and $\psi$.

\begin{figure}[ht] \label{fig:bat2to2dir} \centering
\includegraphics[width=3.2in,height=2.2in]{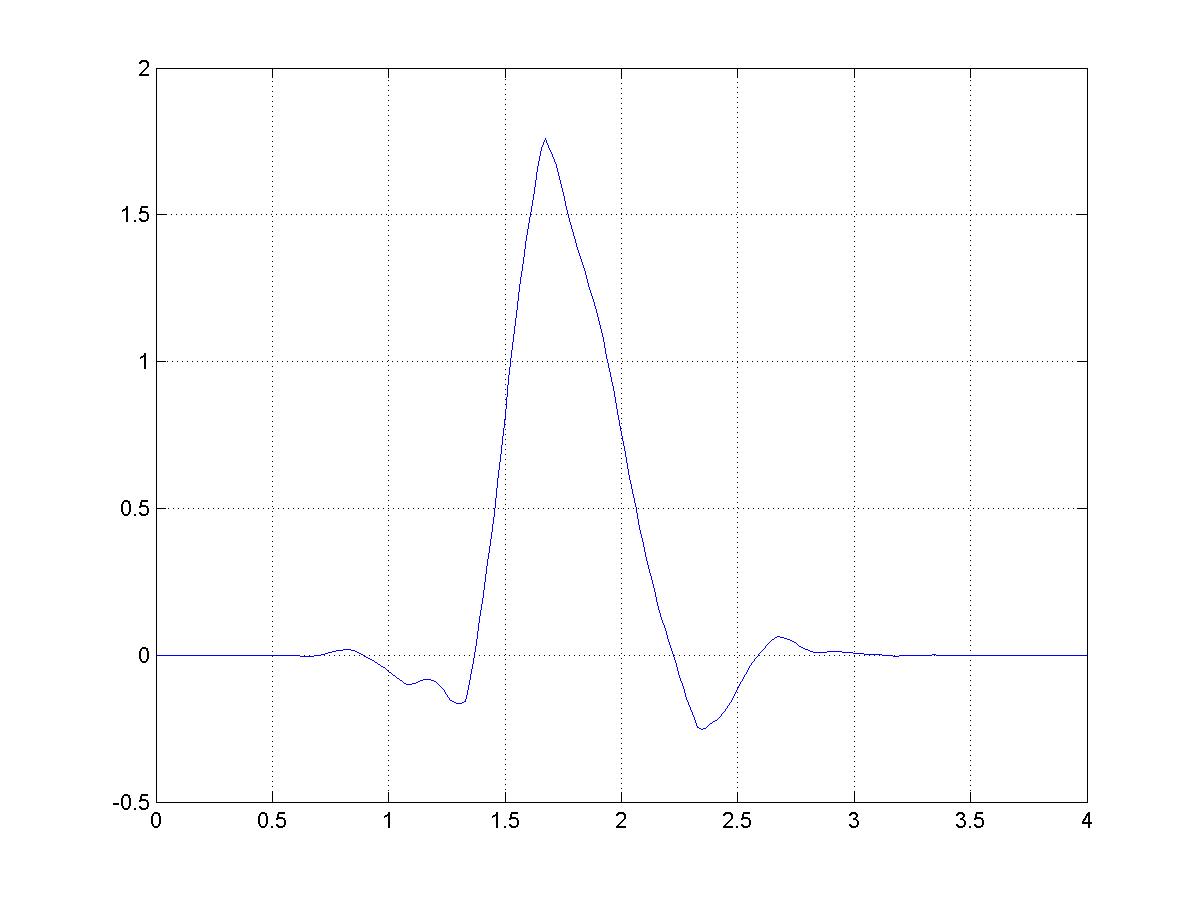} 
\includegraphics[width=3.2in,height=2.2in]{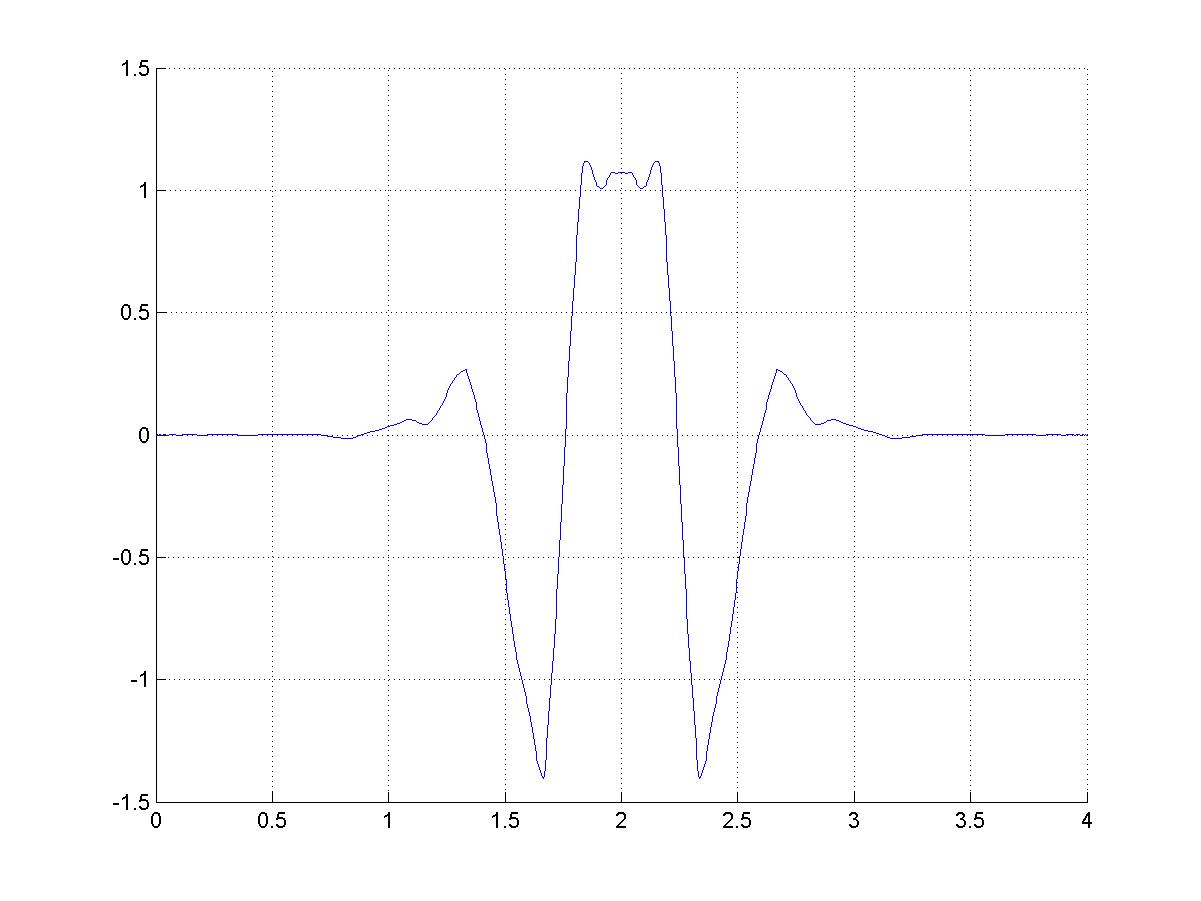} \\
{\title{(a)} \hspace{7.5cm} \title{(b)} } \\
\includegraphics[width=3.2in,height=2.2in]{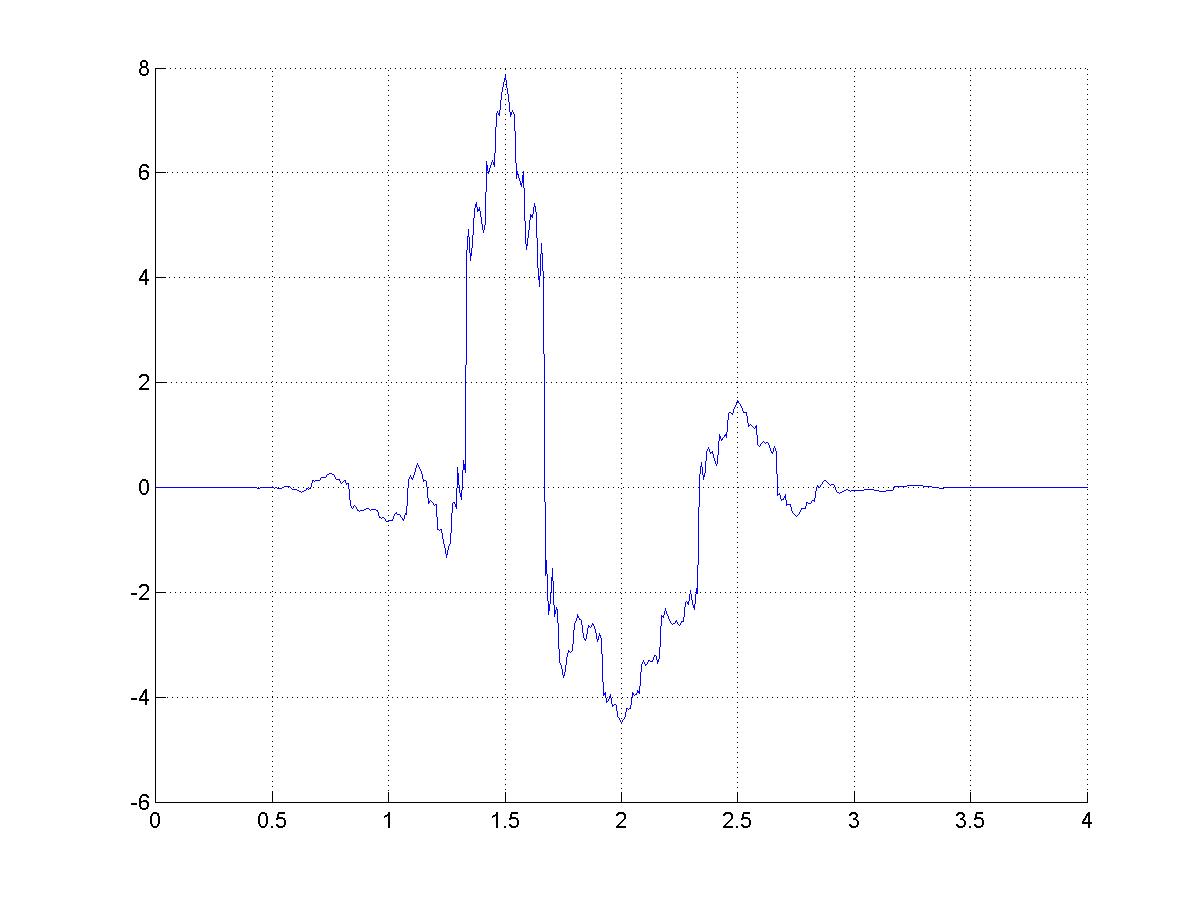} 
\includegraphics[width=3.2in,height=2.2in]{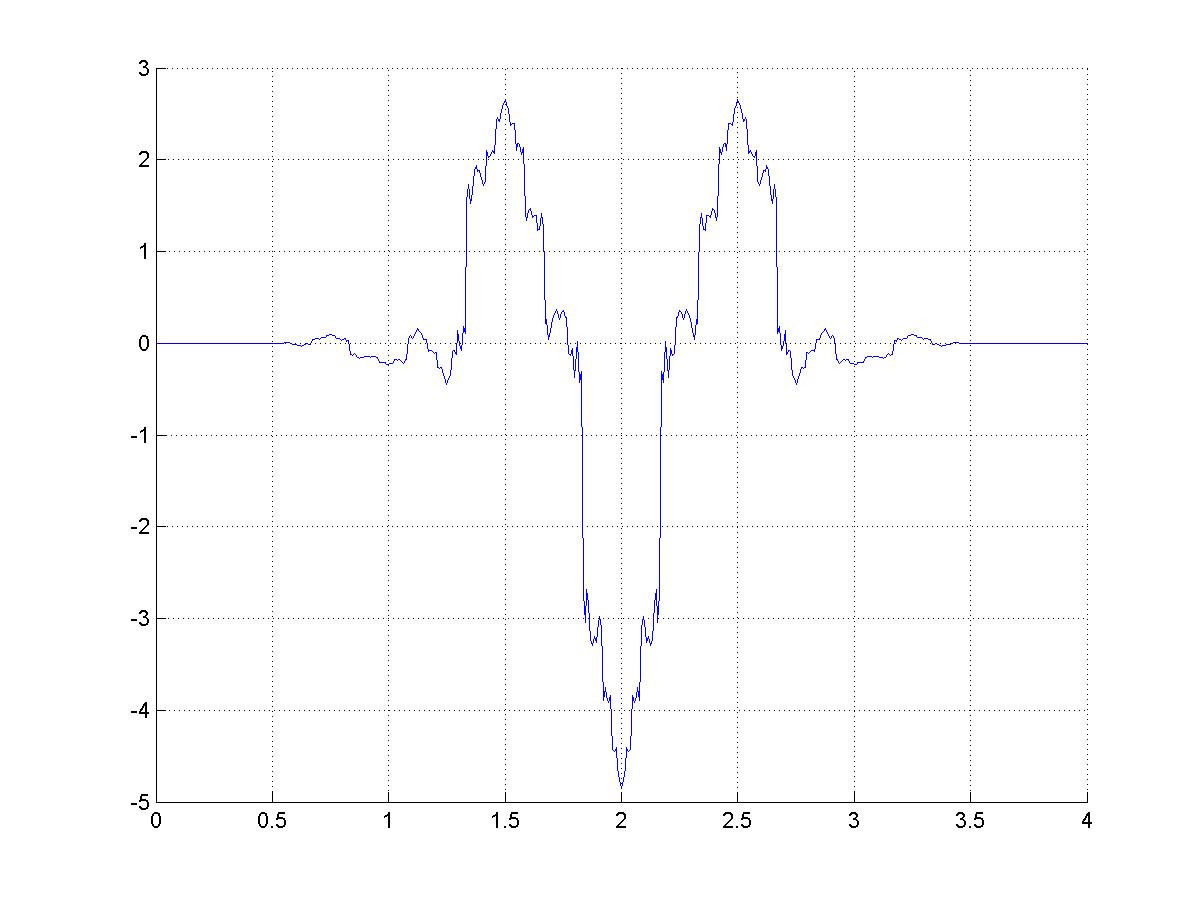} \\
{\title{(c)} \hspace{7.5cm} \title{(d)} }
\caption{
(a) Graph of orthogonal two-direction scaling function $\phi$ of order 2 derived from BAT O2.
(b) Graph of orthogonal two-direction wavelet function $\psi$ derived from BAT O2.
(c) Graph of the first derivative $D\phi$ of $\phi$.
(d) Graph of the first derivative $D\psi$ of $\psi$.
}
\end{figure}

\medskip
{\bf Point values of the first derivative $D\phi$}\\
Since not only $T_\boldphi$ has the eigenvalue $1/2$, but also Sobolev exponent
of $\phi$, 1.6310, is greater than 1.5,
the existence of the first derivative $D\phi$  of $\phi$ is guaranteed.

The eigenvalue problem in~\eqref{eigenphiDn} is
\begin{equation*}
  \begin{bmatrix}
    D\phi(0) \\ D\phi(1) \\ D\phi(2) \\ D\phi(3) \\ D\phi(4)
  \end{bmatrix}
  = d \,T_{D\phi} \begin{bmatrix}
    D\phi(0) \\ D\phi(1) \\ D\phi(2) \\ D\phi(3) \\ D\phi(4)
  \end{bmatrix},
\end{equation*}
where
\begin{equation*}
T_{D\phi} = T_{\phi^{+}} - T_{\phi^{-}}
= \begin{bmatrix}
  0 & 0 & 0 & 0 & -P^{-}_{4} \\
  P^{+}_{2} & P^{+}_{1} & -P^{-}_{4} & -P^{-}_{5} & -P^{-}_{6} \\
  P^{+}_{4}-P^{-}_{4} & P^{+}_{3}-P^{-}_{5} & P^{+}_{2}-P^{-}_{6} & P^{+}_{1}-P^{-}_{7} & 0  \\
  -P^{-}_{6} & -P^{-}_{7} & P^{+}_{4} & P^{+}_{3} & P^{+}_{2} \\
  0 &  0 & 0 & 0 & P^{+}_{4}  \\
\end{bmatrix}.
\end{equation*}
The matrix $T_{D\phi}$ has the eigenvalues $1/2, 0.0116, 0.2359, -0.1480$, and $0$.
The matrix $T_{D\phi}$ has the eigenvector
\begin{equation*}
\begin{bmatrix}
    D\phi(0) & D\phi(1) & D\phi(2) & D\phi(3) & D\phi(4)
\end{bmatrix}^T
\approx \begin{bmatrix}
    0 & 10.1729 & 69.3201 & 1 & 0
\end{bmatrix}^T
\end{equation*}
to the eigenvalue $1/2$.

Since the correct normalization for point values of $D\phi$ is
\begin{equation*}
    \sum_{k\in\Z} k D\phi(k) = - \frac{\sqrt{2}}{2},
\end{equation*}
The normalizing constant is $-10\sqrt{2}/(13+37\sqrt{31}) \approx -0.0646$.
Hence, point values of normalized $D\phi$ at integers are
\begin{equation*}
\begin{bmatrix}
    D\phi(0) & D\phi(1) & D\phi(2) & D\phi(3) & D\phi(4)
\end{bmatrix}^T
\approx \begin{bmatrix}
    0 & -0.6569 & -4.4763 & -0.0646 & 0
\end{bmatrix}^T.
\end{equation*}





\medskip
{\bf Point values of the first derivative $D\psi$}\\
The multiplication in~\eqref{multpsiDn} is
\begin{equation*}
  \begin{bmatrix}
    D\psi(0) \\ D\psi(1) \\ D\psi(2) \\ D\psi(3) \\ D\psi(4)
  \end{bmatrix}
  = d \,T_{D\boldpsi} \begin{bmatrix}
    D\phi(0) \\ D\phi(1) \\ D\phi(2) \\ D\phi(3) \\ D\phi(4)
  \end{bmatrix},
\end{equation*}
where
\begin{equation*}
T_{D\psi} = T_{\boldpsi^{+}} - T_{\boldpsi^{-}}
= \begin{bmatrix}
  0 & 0 & 0 & 0 & -Q^{-}_{4} \\
  Q^{+}_{2} & Q^{+}_{1} & -Q^{-}_{4} & -Q^{-}_{5} & -Q^{-}_{6} \\
  Q^{+}_{4}-Q^{-}_{4} & Q^{+}_{3}-Q^{-}_{5} & Q^{+}_{2}-Q^{-}_{6} & Q^{+}_{1}-Q^{-}_{7} & 0 \\
  -Q^{-}_{6} & -Q^{-}_{7} & Q^{+}_{4} & Q^{+}_{3} & Q^{+}_{2} \\
  0 &  0 & 0 & 0 & Q^{+}_{4}  \\
\end{bmatrix}.
\end{equation*}
Hence, we have
\begin{equation*}
\begin{bmatrix}
    D\psi(0) & D\psi(1) & D\psi(2) & D\psi(3) & D\psi(4)
\end{bmatrix}^T
\approx \begin{bmatrix}
    0 & 0.4775 & 0 & -0.4775 & 0
\end{bmatrix}^T.
\end{equation*}



See Figure ~\ref{fig:bat2to2dir} for the graphs of $D\psi$ and $D\psi$.

\medskip
{\bf Point values of the second derivative $D^{2}\phi$ and $D^{2}\psi$}\\
Since $T_\phi$ does not have eigenvalue $1/4$, we can not find point values of
the second derivatives $D^{2}\phi$ and $D^{2}\psi$.

}
$\hfill\Box$
\end{example} 

\bigskip
\begin{example} \label{ex2} {\rm
In this example we take two-direction multiscaling function $\boldphi$ and two-direction
multiwavelet $\boldpsi$ associated with $\boldphi$ given in ~\citep{WangZhouWang2011}.
We note that multiplicity $r=2$ and dilation factor $d=2$.
%
We shift $P^{+}_k$, $P^{-}_k$, $Q^{+}_k$ and $Q^{-}_k$
so that the supports of $\boldphi$ and $\boldpsi$ are the same,
while orthogonality of $\boldphi$ and $\boldpsi$ is kept.

The nonzero recursion coefficients for $\boldphi$ are
\begin{gather*}
    P^{+}_{1} = \frac{1}{8\sqrt{2}} \begin{bmatrix}
                                     6 & 0 \\ \noalign{\medskip}
                                     -2\sqrt{3}+\sqrt{21} & 3 \\
                                   \end{bmatrix}, \quad
    P^{+}_{2} = \frac{1}{8\sqrt{2}} \begin{bmatrix}
                                     4-2\sqrt{7} & 0 \\ \noalign{\medskip}
                                    3\sqrt{3} & 2-\sqrt{7} \\
                                   \end{bmatrix},
\end{gather*}
\begin{gather*}
    P^{-}_{2} = \frac{1}{8\sqrt{2}} \begin{bmatrix}
                                     4+2\sqrt{7} & 0 \\ \noalign{\medskip}
                                     \sqrt{3} & 2+\sqrt{7} \\
                                   \end{bmatrix}, \quad
    P^{-}_{3} = \frac{1}{8\sqrt{2}} \begin{bmatrix}
                                     2 & 0 \\ \noalign{\medskip}
                                     -2\sqrt{3}-\sqrt{21} & 1 \\
                                   \end{bmatrix}.
\end{gather*}
The nonzero recursion coefficients for $\boldpsi$ are
\begin{gather*}
    Q^{+}_{1} = \frac{1}{8\sqrt{2}} \begin{bmatrix}
                                     0 & -4+2\sqrt{7} \\ \noalign{\medskip}
                                     -2+\sqrt{7} & -3\sqrt{3} \\
                                   \end{bmatrix}, \quad
    Q^{+}_{2} = \frac{1}{8\sqrt{2}} \begin{bmatrix}
                                     0 & 6 \\ \noalign{\medskip}
                                    3 & -2\sqrt{3}+\sqrt{21} \\
                                   \end{bmatrix},
\end{gather*}
\begin{gather*}
    Q^{-}_{2} = \frac{1}{8\sqrt{2}} \begin{bmatrix}
                                     0 & 2 \\ \noalign{\medskip}
                                     1 & -2\sqrt{3}-\sqrt{21} \\
                                   \end{bmatrix}, \quad
    Q^{-}_{3} = \frac{1}{8\sqrt{2}} \begin{bmatrix}
                                     0 & -4-2\sqrt{7} \\ \noalign{\medskip}
                                     -2-\sqrt{7} & -\sqrt{3} \\
                                   \end{bmatrix}.
\end{gather*}
These are shifted and differ from Wang, Zhou, and Wang~\citep{WangZhouWang2011}
by a factor of $1/\sqrt{2}$, due to differences in notation.


Condition E for $\boldphi$ is satisfied.

By ~\eqref{scalarm0} and ~\eqref{mandM22} (detailed computation can be found
in~\citep[Example 4.2]{Kwon2012}), the $j$th continuous moments $\boldm_j$ of $\boldphi$ for $j=0,1,2$ are
\begin{equation*}
\boldm_0 = \frac{\sqrt{2}}{2} \begin{bmatrix}
                   1  \\ \noalign{\medskip}
                   0  \\
                \end{bmatrix},\qquad
\boldm_1 = \frac{7\sqrt{2}-\sqrt{14}}{12} \begin{bmatrix}
                   1  \\ \noalign{\medskip}
                   0  \\
                \end{bmatrix}, \qquad
\boldm_2 = \frac{1}{504}\begin{bmatrix}
                   98(4\sqrt{2}-\sqrt{14}) \\ \noalign{\medskip}
                   6(4\sqrt{6}-\sqrt{42}) \\
                \end{bmatrix}.
\end{equation*}

\medskip
{\bf Point values of $\boldphi$}\\
The eigenvalue problem in~\eqref{eigencpt} is
\begin{equation*}
  \begin{bmatrix}
    \boldphi(0) \\ \boldphi(1) \\  \boldphi(2)
  \end{bmatrix}
  = T_\boldphi \begin{bmatrix}
    \boldphi(0) \\ \boldphi(1) \\ \boldphi(2)
  \end{bmatrix},
\end{equation*}
where
\begin{equation*}
T_\boldphi 
= \frac{1}{\sqrt{2}} \begin{bmatrix}
  \bold{0} & \bold{0} & P^{-}_{2} \\
  P^{+}_{2}+P^{-}_{2} & P^{+}_{1}+P^{-}_{3} & \bold{0} \\
  \bold{0} & \bold{0} & P^{+}_{2} \\
\end{bmatrix},
\end{equation*}
where $\bold{0}$ is the zero matrix of size $2\times 2$.

The eigenvalues of $T_\boldphi$ are 1, 1/2, $-0.0807$, $-0.1614$, and 0 (twice). Hence, we can find point values of $\boldphi$, $\boldpsi$, $D\boldphi$ and $D\boldpsi$, but not $D^2\boldphi$ and $D^2\boldpsi$.

The matrix $T_\boldphi$ has the eigenvector
\begin{equation*}
\begin{split}
&\begin{bmatrix}
    \boldphi(0)^T & \boldphi(1)^T & \boldphi(2)^T
\end{bmatrix}^T
= \begin{bmatrix}
    0 & 0 & -\frac{\sqrt{3}}{3} & 1 & 0 & 0
\end{bmatrix}^T
\end{split}
\end{equation*}
to the eigenvalue 1.

Since the correct normalization for $\boldphi$ is
\begin{equation*}
    \boldm_{0}^{*} \left( \sum_{k\in\Z} \boldphi(k) \right) = \boldm_{0}^{*},
\end{equation*}
the normalizing constant is
$-\sqrt{6}/2 \approx -1.2247$. Hence, point values of normalized $\boldphi$ at integers are
\begin{equation*}
\boldphi(0) = \begin{bmatrix}
            0 \\ \noalign{\medskip}
            0
            \end{bmatrix} , \quad
\boldphi(1) = \begin{bmatrix}
            \frac{\sqrt{2}}{2} \\ \noalign{\medskip}
            -\frac{\sqrt{6}}{2}
            \end{bmatrix}
            \approx \begin{bmatrix}
            0.7071 \\ \noalign{\medskip}
            -1.2247
            \end{bmatrix}, \quad
\boldphi(2) = \begin{bmatrix}
            0 \\ \noalign{\medskip}
            0
            \end{bmatrix}.
\end{equation*}

Once the values of $\boldphi$ at the integers have been determined, we can use
the refinement equation~\eqref{recrelcpt} to obtain values at points of the form
$k/d$, $k \in \Z$, then $k/d^{2}$, and so on to any desired resolution.



{\bf Point values of $\boldpsi$}\\
The multiplication in~\eqref{multpsi} is
\begin{equation*}
  \begin{bmatrix}
    \boldpsi(0) \\ \boldpsi(1) \\ \boldpsi(2)
  \end{bmatrix}
  = T_\psi \begin{bmatrix}
    \boldphi(0) \\ \boldphi(1) \\ \boldphi(2)
  \end{bmatrix},
\end{equation*}
where
\begin{equation*}
T_\boldpsi = \frac{1}{\sqrt{2}} \begin{bmatrix}
  \bold{0} & \bold{0} & Q^{-}_{2} \\
  Q^{+}_{2}+Q^{-}_{2} & Q^{+}_{1}+Q^{-}_{3} & \bold{0} \\
  \bold{0} & \bold{0} & Q^{+}_{2} \\
\end{bmatrix}.
\end{equation*}
Hence, point values of $\boldpsi$ at integers are
\begin{equation*}
\boldpsi(0) = \begin{bmatrix}
            0 \\ \noalign{\medskip}
            0
            \end{bmatrix} , \quad
\boldpsi(1) = \begin{bmatrix}
            \frac{\sqrt{6}}{2} \\ \noalign{\medskip}
            \frac{\sqrt{2}}{2}
            \end{bmatrix}
            \approx \begin{bmatrix}
            1.2247 \\ \noalign{\medskip}
            0.7071
            \end{bmatrix}, \quad
\boldpsi(2) = \begin{bmatrix}
            0 \\ \noalign{\medskip}
            0
            \end{bmatrix}.
\end{equation*}
Once the values of $\boldphi$ and $\boldpsi$ at the integers have been determined,
we can use the equation~\eqref{recrelpsicpt2} to obtain values of
$\boldpsi$ at points of the form $k/d$, $k \in \Z$, then $k/d^{2}$, and so on
to any desired resolution.

See Figure ~\ref{fig:multiex} for the graphs of $\boldphi$ and $\boldpsi$.

%
\begin{figure}[ht] \label{fig:multiex} \centering
\includegraphics[width=3.2in,height=2.2in]{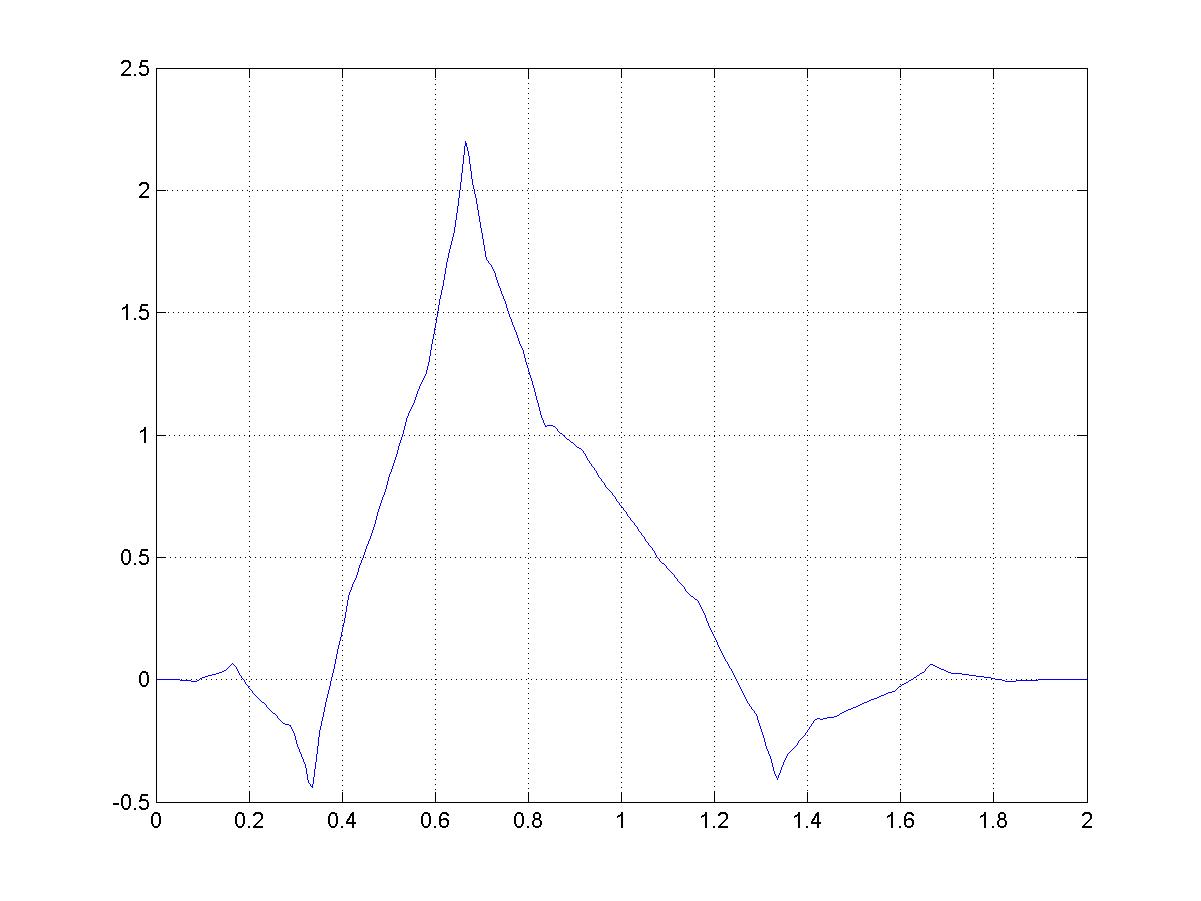} 
\includegraphics[width=3.2in,height=2.2in]{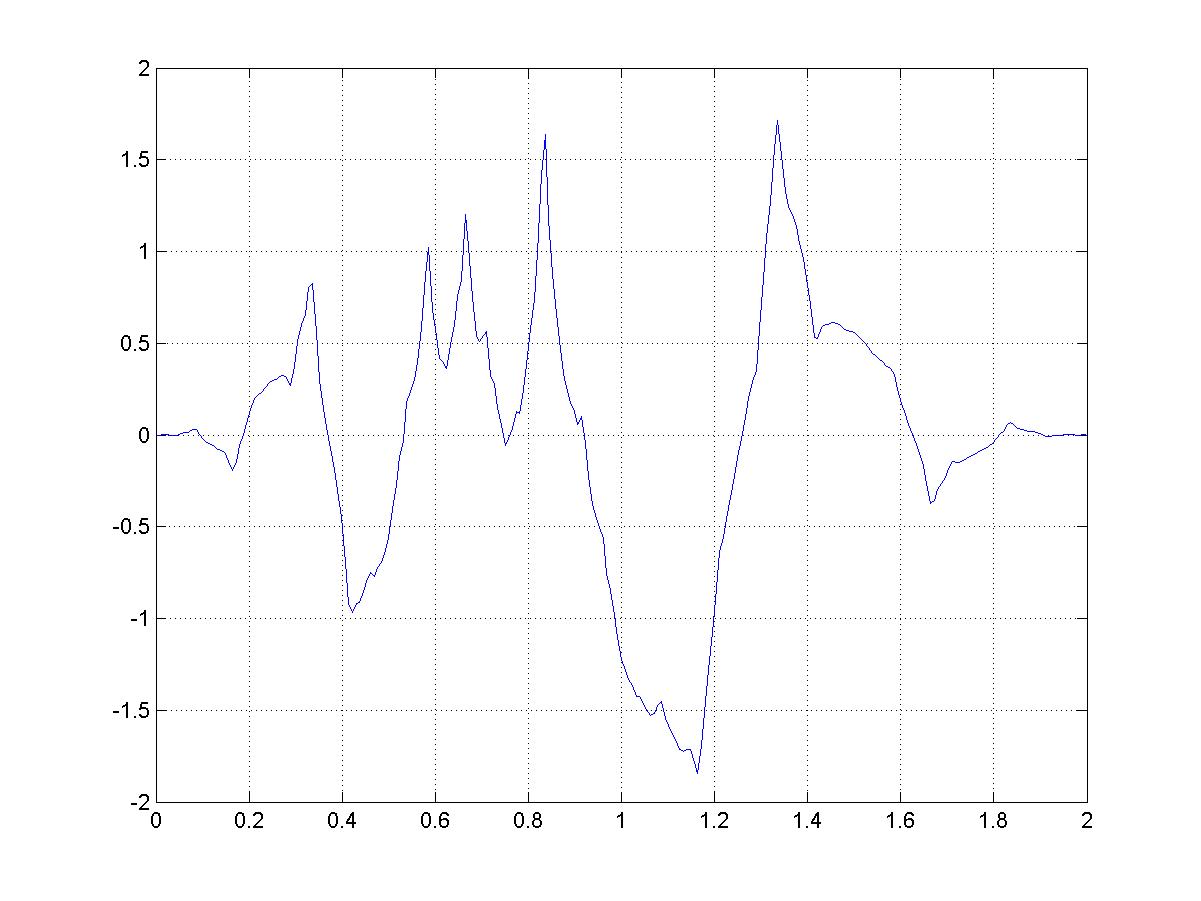} \\
{\title{(a)} \hspace{7.5cm} \title{(b)} }\\
\includegraphics[width=3.2in,height=2.2in]{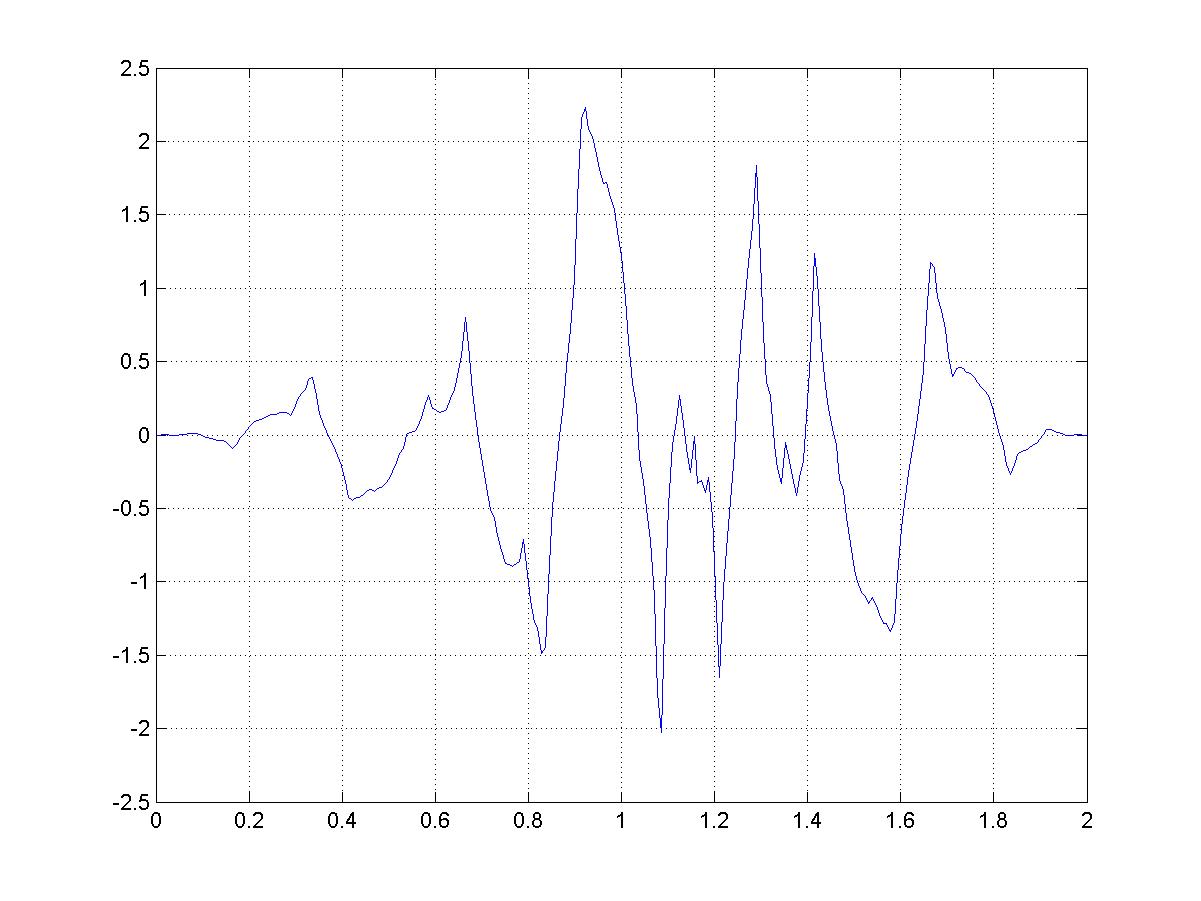} 
\includegraphics[width=3.2in,height=2.2in]{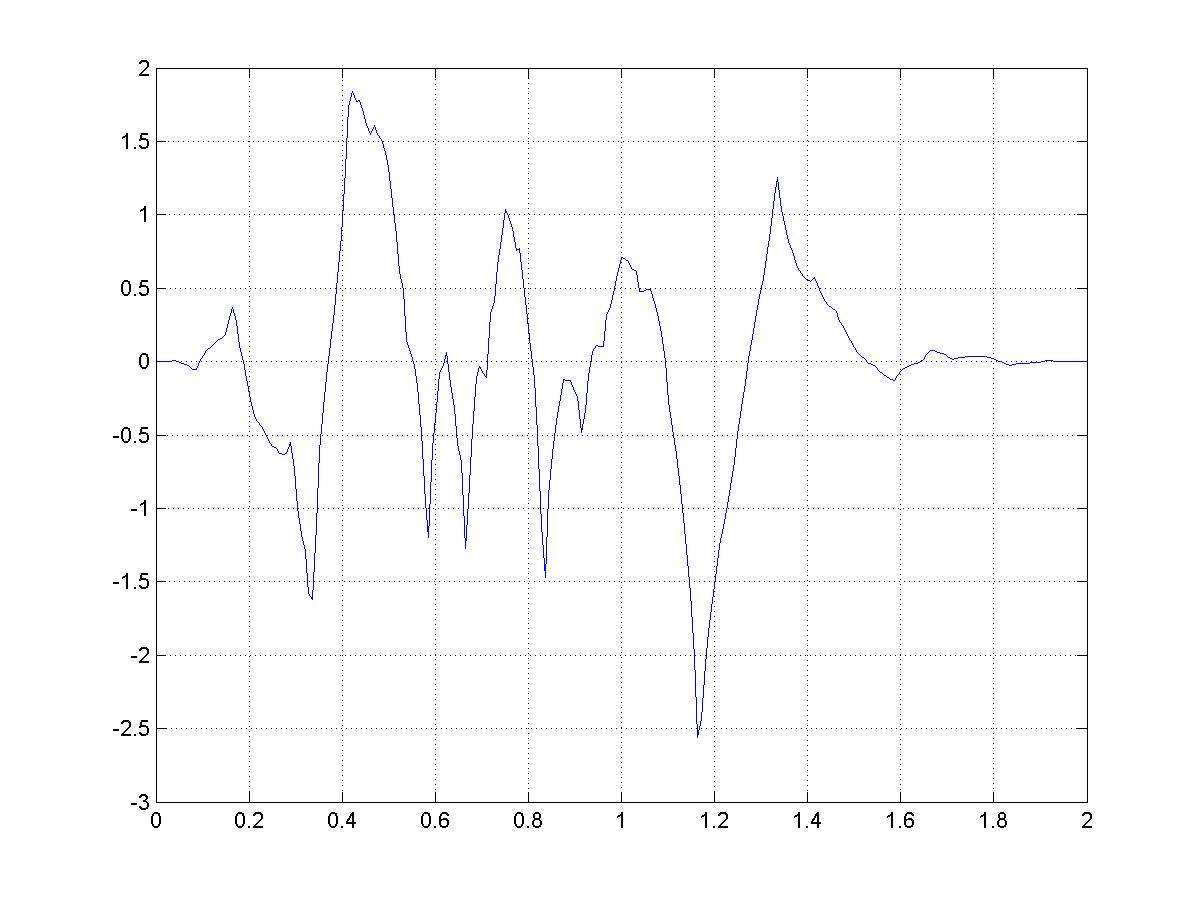} \\
{\title{(c)} \hspace{7.5cm} \title{(d)} }
\caption{
Orthogonal two-direction multiscaling function of order 2 and multiwavelet function:
(a) Graph of $\phi_1$.
(b) Graph of $\phi_2$.
(c) Graph of $\psi_1$.
(d) Graph of $\psi_2$.
}
\end{figure}

However, $T_\boldphi$ has the eigenvalue $1/2$, we do not pursue $D\boldphi$.
The existence of the first derivative $D\boldphi$ of $\boldphi$ is guaranteed
if Sobolev exponent of $\boldphi$ is greater than or equal to 1.5.
Sobolev exponent of $\boldphi$ in this example is 0.5000 and
the existence of $D\boldphi$ is not guaranteed.
}
$\hfill\Box$
\end{example} 




\begin{thebibliography}{10}

\bibitem{daubechies88}
I.~Daubechies.
\newblock Orthonormal bases of compactly supported wavelets.
\newblock {\em Comm. Pure Appl. Math.}, 41(7):909--996, 1988.

\bibitem{daubechies92}
I.~Daubechies.
\newblock {\em Ten lectures on wavelets}, volume~61 of {\em CBMS-NSF Regional
  Conference Series in Applied Mathematics}.
\newblock Society for Industrial and Applied Mathematics (SIAM), Philadelphia,
  PA, 1992.

\bibitem{DuYuan2010}
S.~Du and D.~Yuan.
\newblock The description of two-directional biorthogonal finitely supported
  wavelet packets with poly-scale dilation.
\newblock {\em Trans. Tech. Publications, Switzerland}, 2010.

\bibitem{Keinert2004}
F.~Keinert.
\newblock {\em Wavelets and multiwavelets}.
\newblock Studies in Advanced Mathematics. Chapman \& Hall/CRC, Boca Raton, FL,
  2004.

\bibitem{Kwon2011c}
S.-G. Kwon.
\newblock Approximation order of two-direction multiscaling functions.
\newblock Unpublished results.

\bibitem{Kwon2011d}
S.-G. Kwon.
\newblock High order orthogonal two-direction scaling functions from orthogonal
  balanced multiscaling functions.
\newblock Unpublished results.

\bibitem{Kwon2009}
S.-G. Kwon.
\newblock Characterization of orthonormal high-order balanced multiwavelets in
  terms of moments.
\newblock {\em Bull. Korean Math. Soc.}, 46(1):183--198, 2009.

\bibitem{Kwon2012}
S.-G. Kwon.
\newblock Two-direction multiwavelet moments.
\newblock {\em Appl. Math. Comput.}, 219(8):3530--3540, 2012.

\bibitem{Lebrun-Vetterli-1998b}
J.~Lebrun and M.~Vetterli.
\newblock High order balanced multiwavelets.
\newblock In {\em Proc. IEEE ICASSP, Seatle, WA}, pages 1529--1532, May 1998.

\bibitem{Lebrun-Vetterli-2001}
J.~Lebrun and M.~Vetterli.
\newblock High-order balanced multiwavelets: theory, factorization, and design.
\newblock {\em IEEE Trans. Signal Process.}, 49(9):1918--1930, 2001.

\bibitem{LvWang2009}
B.~Lv and X.~Wang.
\newblock Design and properties of two-direction compactly supported wavelet
  packets with an integer dilation factor.
\newblock In {\em 2009 Third Inter. Symp. on Intel. Info. Tech. Appl.}

\bibitem{Morawiec2009}
J.~Morawiec.
\newblock On ${L}^1$-solution of a two-direction refinable equation.
\newblock {\em J. Math. Anal. Appl}, 354(2):648--656, 2009.

\bibitem{Plonka1998}
G.~Plonka and V.~Strela.
\newblock From wavelets to multiwavelets.
\newblock In {\em Mathematical methods for curves and surfaces, {II}
  ({L}illehammer, 1997)}, Innov. Appl. Math., pages 375--399. Vanderbilt Univ.
  Press, Nashville, TN, 1998.

\bibitem{WangZhouWang2011}
G.~Wang, X.~Zhou, and B.~Wang.
\newblock The construction of orthogonal two-direction multiwavelet from
  orthogonal two-direction wavelet.
\newblock Unpublished results.

\bibitem{XieYang2006}
C.~Xie and S.~Yang.
\newblock Orthogonal two-direction multiscaling functions.
\newblock {\em Front. Math. China}, 1(4):604--611, 2006.

\bibitem{YangLi2007AMC}
S.~Yang and Y.~Li.
\newblock Two-direction refinable functions and two-direction wavelets with
  dilation factor {$m$}.
\newblock {\em Appl. Math. Comput.}, 188(2):1908--1920, 2007.

\bibitem{YangLi2007SciChina}
S.~Yang and Y.~Li.
\newblock Two-direction refinable functions and two-direction wavelets with
  high approximation order and regularity.
\newblock {\em Sci. China Ser. A}, 50(12):1687--1704, 2007.

\bibitem{YangXie2008}
S.~Yang and C.~Xie.
\newblock A class of orthogonal two-direction refinable functions and
  two-direction wavelets.
\newblock {\em Int. J. Wavelets Multiresolut. Inf. Process.}, 6(6):883--894,
  2008.

\end{thebibliography}






\end{document}